\documentstyle[12pt]{article}
\input epsf

\newcommand{\EQ}{\begin{equation}}
\newcommand{\EN}{\end{equation}}
\newcommand{\EQA}{\begin{eqnarray}}
\newcommand{\ENA}{\end{eqnarray}}
\newcommand{\eq}[1]{(\ref{#1})}
\newcommand{\Eq}[1]{equation~(\ref{#1})}
\newcommand{\Eqs}[2]{equations~(\ref{#1}) and~(\ref{#2})}
\newcommand{\Eqss}[2]{equations~(\ref{#1})--(\ref{#2})}
\newcommand{\Sec}[1]{\S\,\ref{#1}}
\newcommand{\Secs}[2]{\S\S\,\ref{#1} and~\ref{#2}}
\newcommand{\Fig}[1]{Figure~\ref{#1}}

\newcommand{\Figs}[2]{Figures~\ref{#1} and \ref{#2}}

\newcommand{\bra}[1]{\langle #1\rangle}
\newcommand{\mean}[1]{\overline #1}
\newcommand{\meanBB}{\overline{\mbox{\boldmath $B$}}}
\newcommand{\meanuu}{\overline{\mbox{\boldmath $u$}}}
\newcommand{\meanuxB}{\overline{\mbox{\boldmath $\delta u\times \delta B$}}}

\newcommand{\xx}{\mbox{\boldmath $x$} {}}

\newcommand{\uu}{\mbox{\boldmath $u$} {}}
\newcommand{\vv}{\mbox{\boldmath $v$} {}}

\newcommand{\bb}{\mbox{\boldmath $b$} {}}

\newcommand{\BB}{\mbox{\boldmath $B$} {}}

\newcommand{\jj}{\mbox{\boldmath $j$} {}}
\newcommand{\JJ}{\mbox{\boldmath $J$} {}}

\newcommand{\EE}{\mbox{\boldmath $E$} {}}

\newcommand{\CC}{\mbox{\boldmath $C$} {}}

\newcommand{\MM}{\mbox{\boldmath $M$} {}}

\newcommand{\nab}{\mbox{\boldmath $\nabla$} {}}

\newcommand{\oo}{\mbox{\boldmath $\omega$} {}}

\newcommand{\emf}{\mbox{\boldmath ${\cal E}$} {}}

\newcommand{\dd}{{\rm d} {}}

\def\la{\mathrel{\mathchoice {\vcenter{\offinterlineskip\halign{\hfil
$\displaystyle##$\hfil\cr<\cr\sim\cr}}}
{\vcenter{\offinterlineskip\halign{\hfil$\textstyle##$\hfil\cr<\cr\sim\cr}}}
{\vcenter{\offinterlineskip\halign{\hfil$\scriptstyle##$\hfil\cr<\cr\sim\cr}}}
{\vcenter{\offinterlineskip\halign{\hfil$\scriptscriptstyle##$\hfil\cr<\cr\sim\cr}}}}}

\newcommand{\ea}{{\em et al. }}

\newcommand{\yan}[5]{, ``#5,'' {\em Astron. Nachr. }{\bf #2}, #3-#4 (#1).}

\newcommand{\yana}[5]{, ``#5,'' {\em Astron. Astrophys. }{\bf #2}, #3-#4 (#1).}

\newcommand{\ysph}[5]{, ``#5,'' {\em Solar Phys. }{\bf #2}, #3-#4 (#1).}

\newcommand{\ymn}[5]{, ``#5,'' {\em Monthly Notices Roy. Astron. Soc. }
{\bf #2}, #3-#4 (#1).}

\newcommand{\ynat}[5]{, ``#5,'' {\em Nature }{\bf #2}, #3-#4 (#1).}

\newcommand{\yjfm}[5]{, ``#5,'' {\em J. Fluid Mech. }{\bf #2}, #3-#4 (#1).}
\newcommand{\ypr}[5]{, ``#5,'' {\em Phys. Rev. }{\bf #2}, #3-#4 (#1).}
\newcommand{\yprl}[5]{, ``#5,'' {\em Phys. Rev. Lett. }{\bf #2}, #3-#4 (#1).}

\newcommand{\yapj}[5]{, ``#5,'' {\em Astrophys. J. }{\bf #2}, #3-#4 (#1).}

\newcommand{\yapjl}[5]{, ``#5,'' {\em Astrophys. J. Lett. }
{\bf #2}, #3-#4 (#1).}

\newcommand{\ypf}[5]{, ``#5,'' {\em Phys. Fluid }
{\bf #2}, #3-#4 (#1).}

\newcommand{\ygafd}[5]{, ``#5,'' {\em Geophys. Astrophys. Fluid Dynam.}
{\bf #2}, #3-#4 (#1).}

\newcommand{\yjour}[6]{, ``#6,'' {\em #2} {\bf #3}, #4-#5 (#1).}

\newcommand{\yproc}[7]{, ``#4,'' In {\em #5} (ed. #6), pp. #2-#3. #7 (#1).}
\newcommand{\ybook}[3]{ {\em #2}. #3 (#1).}

\newcommand{\pgafd}[2]{ ~#1~ ``#2,'' {\em Geophys. Astrophys. Fluid Dynam. } (in press)}

\begin{document}
\begin{center}
\LARGE
{\uppercase{\bf Local and nonlocal magnetic diffusion and alpha-effect
tensors in shear flow turbulence}}
\Large
\vspace{1em}

\uppercase{Axel Brandenburg}\footnote{Now at NORDITA, Blegdamsvej 17,
DK-2100 Copenhagen \O, Denmark} \& \uppercase{Dmitry Sokoloff}\footnote{on
leave of absence from: Department of Physics, Moscow State University,
119899 Moscow, Russia}
\vspace{1em}

\normalsize
{\it Department of Mathematics, University of Newcastle upon Tyne, NE1 7RU, UK}

\tiny(\today~ $ $Revision: 1.27 $ $)\normalsize
\end{center}

\begin{abstract}
Various approaches to estimate turbulent transport coefficients from
numerical simulations of hydromagnetic turbulence are discussed.  A
quantitative comparison between the averaged magnetic field obtained
from a specific three-dimensional simulation of a rotating turbulent
shear flow in a slab and a simple one-dimensional alpha-omega dynamo
model is given. A direct determination of transport coefficients is
attempted by calculating the correlation matrix of different components
of the field and its derivatives.  This matrix relates the
electromotive force to physically relevant parameters like the tensor
components of the alpha-effect and the turbulent diffusivity. The
alpha-effect operating on the toroidal field is found to be negative
and of similar magnitude as the value obtained in previous work by
correlating the electromotive force with the mean magnetic field. The
turbulent diffusion of the toroidal field is comparable to the
kinematic viscosity that was determined earlier by comparing the stress
with the shear. However, the turbulent diffusion of the radial field
component is smaller and can even be formally negative.  The method is
then modified to obtain the spectral dependence of the turbulent
transport coefficients on the wavenumber. There is evidence for
nonlocal behavior in that most of the response comes from the smallest
wavenumbers corresponding to the largest scale possible in the
simulation. Again, the turbulent diffusion coefficient for the radial
field component is small, or even negative, which is considered
unphysical. However, when the diffusion tensor is assumed to be diagonal
the radial component of the diffusion tensor is positive, supporting
thus the relevance of a nonlocal approach. Finally, model calculations
are presented using nonlocal prescriptions of the $\alpha$-effect and
the turbulent diffusion. We emphasize that in all cases the electromotive
force  exhibits a strong stochastic component which make the alpha-effect
and the turbulent diffusion intrinsically noisy.
\end{abstract}

\section{Introduction}

In recent years, significant progress has been made in understanding
the generation of large scale magnetic fields by dynamo action. One of
the outstanding questions is whether such large scale dynamos work in
the limit of high conductivity and in the presence of finite amplitude
magnetic fields. These questions have so far mostly been addressed in
the framework of fully periodic boxes with external forcing (Cattaneo \&
Vainshtein 1991, Vainshtein \& Cattaneo 1992, Cattaneo \& Hughes 1996,
Brandenburg 2001; hereafter referred to as B01).
Subsequently, concerns have come up, because periodic
boxes have the peculiar property that magnetic helicity is conserved. In
that case the $\alpha$-effect may well be `catastrophically'
quenched by finite amplitude magnetic fields if the magnetic Reynolds
number is large (Blackman \& Field 2000, B01).

Unfortunately, externally forced simulations with non-periodic,
open boundaries do not seem to alleviate the catastrophic quenching
problem (Brandenburg \& Dobler 2001; hereafter referred to as BD). This is the
reason why there is now an urgent need to consider simulations of naturally driven
turbulence. Apart from convection, for which the $\alpha$-effect has
been studied extensively (Brandenburg \ea 1990, Ossendrijver \ea 2001),
shear flow turbulence that is driven by the magnetorotational instability
is another important example (Brandenburg et al.\ 1995, hereafter referred
to as BNST95, Hawley et al.\ 1996). In the case of naturally driven
turbulence, $\alpha$-effect and turbulent diffusivity
are no longer scalars but tensors, nor
are they just coefficients but integral kernels. The purpose of this
paper is to explore these aspects of inhomogeneous turbulence using
three-dimensional simulations.

\begin{figure}[t!]
\epsfbox{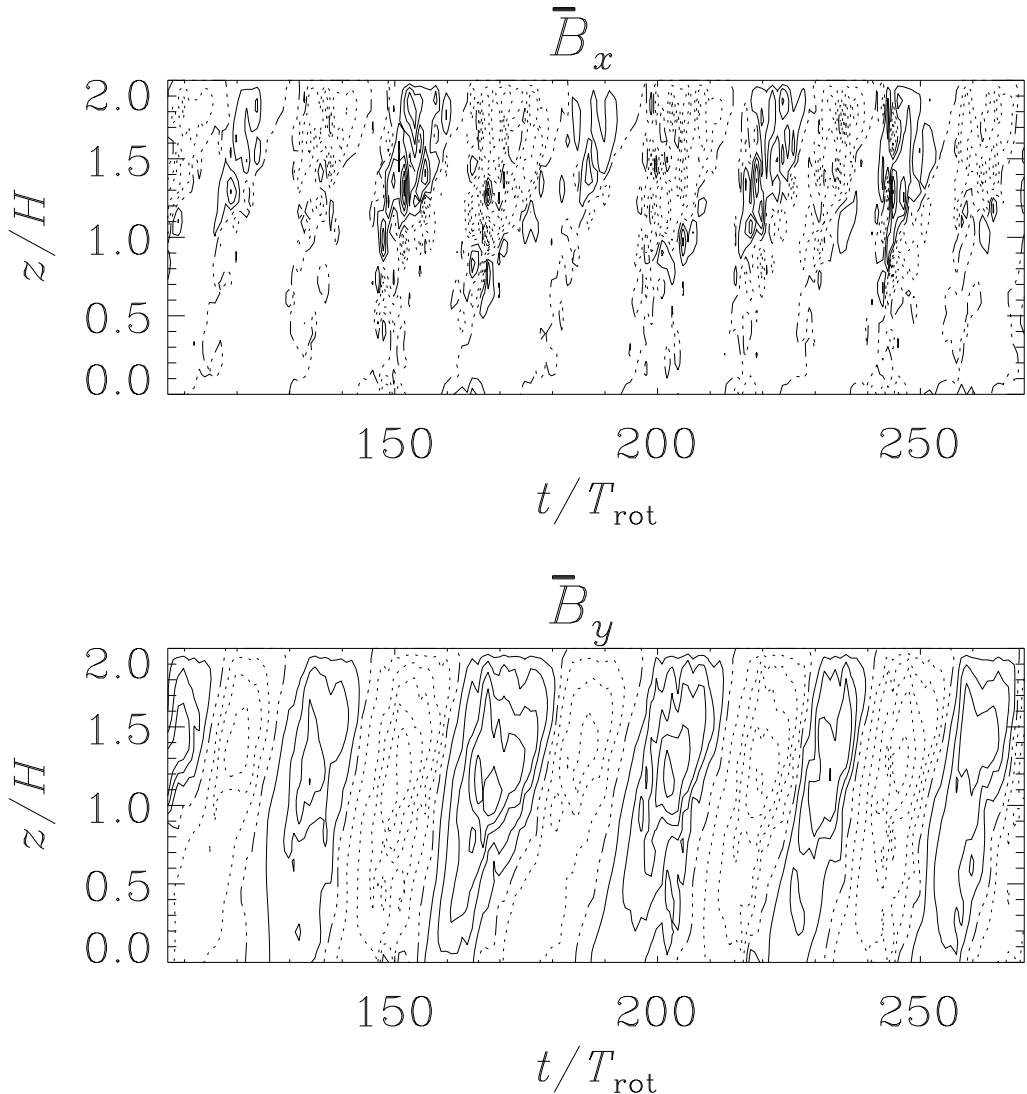}
\caption[]{Horizontally averaged radial and toroidal magnetic fields,
$\mean{B_x}$ and $\mean{B_y}$ respectively, as a function of time
and height, as obtained from the fully three-dimensional simulation
of BNST96. Time is given in units of rotational periods, $T_{\rm rot}$.
(No smoothing in $z$ or $t$ is applied.)
Dotted contours denote negative values.
}\label{Fbutter}\end{figure}

Simulations relevant to astrophysical
bodies or to the Earth that include the effects of
rotation and large scale velocity shear have recently displayed
strikingly coherent spatio-temporal order (Glatzmaier \& Roberts 1995,
BNST95). As an
example we reproduce in \Fig{Fbutter} a space-time diagram (or
butterfly diagram in solar physics) of the mean magnetic field of an
accretion disc simulation of Brandenburg et al.\ 1996 (hereafter
referred to as BNST96), which is an extended run of BNST95. In this
particular simulation (Run~0 of BNST96) the symmetry of the magnetic
field has been restricted to even parity, so the computation has been
carried out in the upper disc plane, $0<z<L_z$, where $L_z=2H$ is the
vertical extent of the box, $H$ is the gaussian scale height of the
hydrostatic equilibrium density, and $z=0$ corresponds to the
equatorial plane.

\epsfxsize=16cm

It is interesting to note that the spatio-temporal behavior obtained
from the three-dimensional simulations resembles in many ways what
has been obtained earlier using mean-field models. In particular, the
magnetic fields of the simulations of Glatzmaier \& Roberts (1995)
show magnetic field patterns with basic dipole symmetry and occasional
field reversal, similar to what is seen in mean-field dynamo models
(e.g.\ Hollerbach et al.\ 1992, Jones \& Wallace 1992). Of course, the
simulations of Glatzmaier \& Roberts (1995) are much more realistic and
show additional features such as the magnetic dipole inclination.
However, we are not aware of a detailed attempt to relate the
three-dimensional simulations to mean-field theory. At present there
are no estimates of the relevant alpha-effect, or even its sign,
and whether it is actually the alpha-effect that is responsible
for the large scale field generation in these simulations.
Other proposals for causing large-scale field generation include,
for example, the incoherent $\alpha$-effect (Vishniac \& Brandenburg
1997), negative magnetic diffusivity effects (Zheligovsky, Podvigina,
\& Frisch 2001), or effects that conserve magnetic helicity explicitly
(Vishniac \& Cho 2001, although there is non-supportive evidence;
see Arlt \& Brandenburg 2001).

The aim of this paper is to assess the possible relation between
simulations and mean-field theory using a local simulation of BNST96
that is readily available to us. This simulation has the advantage
of being in cartesian geometry, and one can define averages that are
naturally dependent on just one spatial coordinate ($z$). Furthermore,
geometrical or boundary effects are not very pronounced in that model.

In the particular process considered by BNST95 an initial magnetic
field of suitable magnitude, but vanishing average
($\mean{B_x}=\mean{B_y}=\mean{B_z}=0$), leads after at least thirty
rotational periods to a self-sustained turbulent state, which is
statistically steady and independent of the detailed initial
conditions. This process results from two instabilities, both with
positive feedback. There is the magnetic shearing or Balbus-Hawley
(1991) instability that generates the turbulence and a dynamo
instability that regenerates the magnetic field. The energy that drives
the dynamo is constantly being tapped from the large scale velocity
shear.

At first glance it may seem inappropriate to invoke a mean-field
description in the present case which is highly nonlinear and where the
turbulence itself is driven by the dynamo-generated magnetic field.
However, we adopt here the view that the mechanism that leads to the
generation of large scale magnetic fields is distinct from the
feedback cycle that leads to the smaller scale magnetic fields
driving the turbulence, which in turn drives the magnetic field.
There have been some attempts to model this feedback cycle using
a set of phenomenological equations (Tout \& Pringle 1992, Reg\"os 1997),
but these models too involve an $\alpha$-effect that is prescribed
in an {\it ad hoc} fashion.
Unfortunately, these models do not reproduce a number of
features seen in the simulations (ratio of poloidal to toroidal fields,
magnitude of the accretion torque, role of the Parker instability, and
significance of the vertical field.) Also, it should be emphasized that
the large scale field seen in BNST95 is really a consequence of
vertical stratification.  Models without stratification (Hawley et
al.\ 1996) do not show large scale fields, although the feedback cycle
of dynamo-generated turbulence still works. This confirms that the
large scale field generation can indeed be considered as a process
operating on top of the otherwise highly nonlinear feedback cycle
leading to dynamo-generated turbulence. If this is the case one might
be able to describe the large scale dynamics in terms of averaged
equations. This would imply a closure relation between the nonlinear
induction term (the electromotive force) and the mean magnetic field.
It is this relation which is at the center of the present paper.

On the one hand, our findings support basic aspects of the mean-field
concept. On the other hand, when trying to estimate the mean-field
transport coefficients (alpha-effect and turbulent diffusion) it
becomes necessary to go beyond simple parameterizations, to adopt
tensorial and nonlocal forms of closures, and to allow for stochastic
effects. While these results are primarily applicable to local simulations
of accretion discs, it will be interesting to make qualitative comparisons
with simulations of other systems (geodynamo, solar dynamo).

Our parameterization of transport coefficients depends implicitly on
the magnetic field strength, because the data used are from a fully
nonlinear calculation. However, we are not able to investigate this
dependence in our present approach, because the time-averaged level of
magnetic field is rather stable in the model.
We should however emphasize that in the present model, where
the turbulence is magnetically driven, the magnetic energy is strong
(three times in superequipartition with the turbulent energy) and that
the turbulent transport coefficients are therefore already strongly
affected, and possibly even enhanced, by this strong magnetic field.

Before we begin we summarize a few important properties of the
simulation of BNST96. The calculation was carried out in a cartesian
box where overall rotation, radial linear shear and vertical density
stratification are included. The boundaries in the toroidal ($y$)
direction are periodic and in the radial ($x$) direction sliding
periodic (Hawley et al.\ 1995), which is periodic with respect to
positions that shift in time. At the top the boundaries are
impenetrable and stress free, and the horizontal components of the
magnetic field vanish. At the lower boundary of the box a symmetry
condition is applied.

We consider the horizontally averaged magnetic field components in the
two horizontal directions, $\mean{B_x}$ and $\mean{B_y}$; see \Fig{Fbutter}.
The averaged vertical field is conserved, because periodic (or
shearing-periodic) boundary conditions are used in the horizontal directions. 
Since the averaged
vertical field vanishes initially, it vanishes at all times, see BNST95.
We observe the following remarkable properties of the horizontally
averaged field:
(i) its sign changes every approximately 15 rotational periods,
(ii) at larger heights the reversals occur somewhat later than at
lower heights, and
(iii) the toroidal and radial fields are out of phase by about $3\pi/4$.
The latter is more clearly seen in curves showing the averaged field
as a function of time (e.g.\ Fig.~1c in BNST96).
Property (ii) can be interpreted as a pattern migration away from the
equatorial plane at speed $c>0$ ($c\approx0.024\Omega H$). In addition,
there are properties related to the symmetry of the field about the
equatorial plane; see Brandenburg (1998) for a discussion of the
dependence on boundary conditions (and the striking agreement with
predictions from simple $\alpha\Omega$-dynamos). 

We begin by discussing the signs and magnitudes of various
helicities that are relevant in connection with the $\alpha$-effect.
We then compare quantitatively a simple mean field model with
the result of the simulations. We also consider the direct
determination of turbulent transport coefficients allowing for the
possibility that the transport coefficients may depend either on $z$
or on the vertical scale (i.e.\ the vertical wavenumber).
Finally we present a model with a scale dependent (or nonlocal)
formulation of magnetic diffusion and $\alpha$-effect, as well as
a model where strong stochastic fluctuations are included.

\section{Magnetic and kinetic helicities}
\label{Shelicities}

In the light of recent results for $\alpha$ and $\eta_{\rm t}$ in
forced turbulence, it is useful to discuss some relevant
properties of the present accretion disc simulations. Firstly, as
expected for a rotating stratified medium, the kinetic helicity is
negative in the upper disc plane. In B01 the magnitude of the ratio
$k_{\rm eff}\equiv\overline{\oo\cdot\uu}/\overline{\uu^2}$ was found to be equal to the
forcing wavenumber, $k_{\rm f}$. (Here, $\oo=\nabla\times\uu$ is vorticity and
overbars denote horizontal averaging.) Applying
this to the disc simulations we find the {\it effective} forcing wavenumber
between 2 and 4 in $H\leq z\leq2H$, and nearly zero in $0\leq z\leq H$;
see \Fig{Fbutter}.
These wavenumbers correspond to a length scale of about 3, which is comparable
to the scale of the box. Thus, one cannot expect to see some of the
pronounced features of B01 associated with an at least modest amount of
scale separation.

\begin{figure}[t!]
\epsfbox{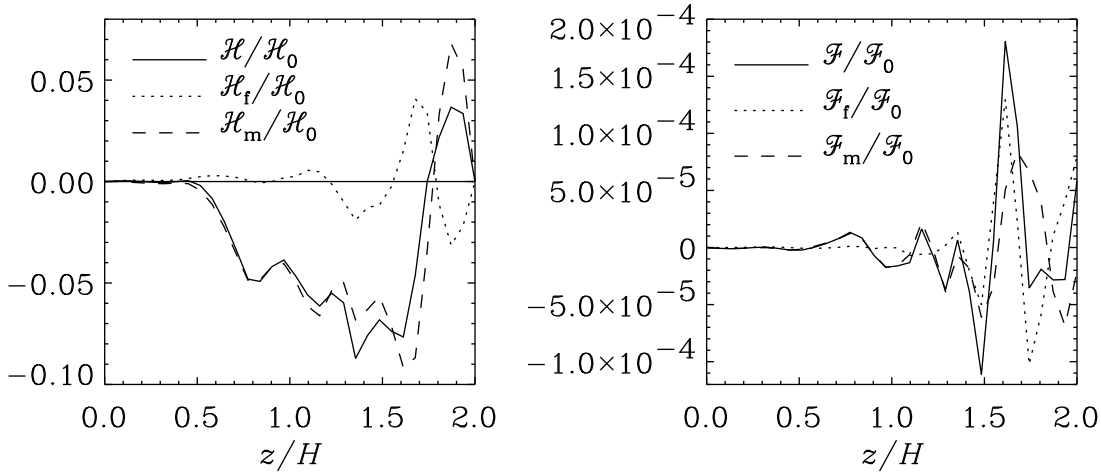}
\caption[]{Vertical dependence of magnetic helicity
${\cal H}={\cal H}_{\rm m}+{\cal H}_f$ and magnetic helicity flux
${\cal F}={\cal F}_{\rm m}+{\cal F}_f$, where the subscripts m and f
denote the contributions from the mean and fluctuating fields. Magnetic
helicity is normalized by ${\cal H}_0=k_1^{-1}\int\BB^2\dd V$, where
$k_1=\pi/L_z$ is the smallest wavenumber in the vertical direction. The
magnetic helicity flux is normalized by ${\cal F}_0={\cal H}_0\Omega H$,
where $H=L_z/2$ is the density scale height of the disc.
}\label{Fpabm}\end{figure}

In B01 the small scale current helicity, $\overline{\jj\cdot\bb}$,
was of the same sign as $\overline{\oo\cdot\uu}$ and of the same
magnitude.  (Here, $\jj=\nabla\times\bb$ is the small scale
current density in units where the magnetic permeability is unity.)
This is not the case in the present simulations. First
of all, $\overline{\jj\cdot\bb}$ is positive near the top of
the box in $1.5H\leq z\leq2H$, and at most only about 10\% of
$|\overline{\oo\cdot\uu}|$. In $0\leq z\leq1.5H$ the small scale
current helicity is essentially fluctuating about zero. However, as
in B01, there is also here a tendency for the total current helicity,
$\overline{\JJ}\cdot\overline{\BB}+\overline{\jj\cdot\bb}$, to cancel
to zero, although not quite, and only near the top boundary where large
scale and small scale current helicities tend to have opposite signs. This
is however different from the model studied by BD,
where open boundary conditions (the
same that are used here) were found to yield a modest helicity flux out
of the domain, which offsets the otherwise strict steady state balance
between large scale and small scale current helicities that was found
in B01 and also here. In the present simulations the magnetic helicity
flux, ${\cal F}$, is however very small; see \Fig{Fpabm}.

It is somewhat surprising that in the present simulations the resulting
magnetic helicity flux is so weak. This could be related to the fact that
here, in contrast to BD, the turbulence is driven self-consistently
by the resulting magnetic field. Near the boundaries there are relatively sharp
gradients in all quantities, which is typical of boundary layer behavior,
and is unlikely to occur in more realistic cases where free turbulent
exchange and flows through the boundaries can occur.

\epsfxsize=16cm

The fact that the small scale current helicity is positive indicates
that $\alpha$ is negative. The connection between the two
is based on a formula given by Keinigs
(1983), $\alpha=-\eta\,\overline{\jj\cdot\bb}/\overline{\BB}^2$, where
$\eta$ is the microscopic magnetic diffusivity.\footnote{We note that the
results of B01 are compatible with a version of Keinigs' formula where
$\alpha-\eta_{\rm t}\,\overline{\JJ}\cdot\overline{\BB}/\overline{\BB}^2$
is equal to $-\eta\,\overline{\jj\cdot\bb}/\overline{\BB}^2$; see Eq.~(48)
of B01. However, for the dynamo to be excited, the $\eta_{\rm t}$ term
must always be slightly smaller than the $\alpha$ term. Therefore the
sign implied for $\alpha$ is not affected by this generalization.}
Another more direct estimate
for $\alpha$ comes from considering the balance of different induction
effects, $\alpha\overline{\BB}+\overline{\uu}\times\overline{\BB}$
on the one side and $(\eta+\eta_{\rm t})\overline{\JJ}$ on the other.
Taking the dot product with $\overline{\BB}$ shows that
$\alpha=+\eta_{\rm t}\overline{\JJ}\cdot\overline{\BB}/\overline{\BB}^2$.
Thus, since $\overline{\JJ}\cdot\overline{\BB}$ is found to be negative
we again find that $\alpha$ is negative in the upper disc plane.
(The negative sign of $\overline{\JJ}\cdot\overline{\BB}$ is in agreement
with a negative sign of the magnetic helicity ${\cal H}$ throughout most
of the computational domain; see \Fig{Fpabm}.)

Perhaps the most convincing explanation for the negative $\alpha$ is that
intense parts of a flux tube contract (to maintain pressure balance along
field lines), but are also most buoyant. If this contraction is stronger
than the expansion associated with the rise into a less dense medium,
then $\alpha$ will be negative (Brandenburg 1998). The same result was
also obtained by R\"udiger \& Pipin (2000) under the assumption that the
turbulence is driven primarily by small scale magnetic fields.

Next we compare quantitatively the results of the simulations with a
simple mean field model. Here the negative sign of $\alpha$ follows
directly from the fact that the dynamo wave is found to migrate away
from the disc plane.

\section{Comparison with conventional mean field dynamos}
\label{Scomparison}

In mean-field dynamo theory one uses the averaged induction equation,
\EQ
{\partial\meanBB\over\partial t}=\nab\times(\meanuu\times\meanBB+\meanuxB),
\EN
where overbars denote horizontal averages and the deltas denote
fluctuations. Our horizontal averages may be associated with ensemble
averages in analytical studies.
In astrophysical bodies with large scale shear, $\meanuu=\meanuu(\xx)$,
e.g.\ differential rotation, a mean component of the magnetic field
is generated in the direction of the shear.
However, this only works as long as there is a mechanism replenishing
the magnetic field component in the cross stream direction.
Such a field could arise from the correlation term $\meanuxB$, the
``turbulent'' electromotive force, which we denote by $\emf$.
Early work since the 1960s and 1970s has established the
following form for $\emf$:
\EQ
{\cal E}_i = \alpha_{ij} \mean{B_j}
+\eta_{ijk}\partial\mean{B_j}/\partial x_k
\label{emf}
\EN
(Roberts \& Soward 1975, Krause \& R\"adler 1980, see also Steenbeck,
Krause \& R\"adler 1966 for an early reference, and Parker 1955 for the
original formulation of the idea).  In this representation one assumes
that the averaged magnetic field $\meanBB$ is smooth so that higher
derivatives in the Taylor expansion \eq{emf} do not enter. It is then
assumed that the functional relationship between $\emf$ and $\meanBB$
is local. We return to the applicability of a local relationship in the
next section.

An apparently less general formulation of the diffusion term in \Eq{emf} is
\EQ
{\cal E}_i=\alpha_{ij}\mean{B_j}-\eta^*_{ij}\mean{J_j},
\label{emf2}
\EN
where diffusion works only via the current density $\JJ=\nab\times\BB$
(in units where the magnetic permeability is unity).
However, the two cases are here equivalent, because in \Eq{emf} only the index $k=z$
gives a nonvanishing contribution (horizontal derivatives of horizontal
averages vanish in our case!). Therefore the matrices $\eta_{ijz}$ and
$\eta^*_{ij}$ contain the same information. In particular,
\EQ
\pmatrix{\eta^*_{xx}&\eta^*_{xy}\cr\eta^*_{yx}&\eta^*_{yy}}=
\pmatrix{\eta_{xyz}&-\eta_{xxz}\cr\eta_{yyz}&-\eta_{yxz}}.
\label{relation}
\EN
The advantage of formulation \eq{emf2} is that it is straightforward to ensure
that turbulent diffusion does indeed lead to a {\it decrease} of magnetic
energy. This is the case when the matrix $\eta^*_{ij}$ is positive
definite, i.e.\ $\eta^*_{ij}J_iJ_j>0$. A necessary condition for this is
$\eta^*_{xx}>0$ and $\eta^*_{yy}>0$. Of particular interest
is the case where $\eta^*_{ij}$ is diagonal, i.e.\ $\eta^*_{xy}$ and
$\eta^*_{yx}$ are {\it assumed} to vanish. Below we shall consider both
representations, \eq{emf} and \eq{emf2} with $\eta^*_{xy}=\eta^*_{yx}=0$,
and correspondingly we use
different notation in the two cases, $\eta_{ijz}$ and $\eta^*_{ij}$,
respectively. We emphasize that \Eq{relation} does not hold if the
off-diagonal components of $\eta^*_{ij}$ are put to zero, whilst
$\eta_{ijz}$ is allowed to have all four components different from
zero.

In principle there could be an additional gradient term in the expression
for $\cal E$ that would not affect the evolution of
the field, but would modify the fitting procedure for the transport
coefficient. (We are grateful to N.~Kleeorin and I.~Rogachevskii, who
attracted our attention to this possibility.) In the present case where
averages depend only on $z$, such gradient terms correspond
simply to a constant vector, $\EE_0$; see, e.g., BD. However,
in the present case the horizontal components of $\emf$ are
antisymmetric about the midplane and have to vanish there. Since
the terms on the right hand side of \Eqs{emf}{emf2} are also zero
on $z=0$ we must have $\EE_0=0$.

Now in order that the poloidal field can be replenished it is important
to have a nonvanishing component of $\alpha_{yy}$. The magnitude
of $\alpha_{yy}$ has to be sufficiently large to overcome the effects
of (turbulent) diffusion that arise from $\eta_{\rm t}\epsilon_{ijk}$,
the isotropic part of $\eta_{ijk}$.

In BNST95, and later in Brandenburg \& Donner 1997, the value of
$\alpha_{yy}$ was roughly estimated in the following way. It turned
out that ${\cal E}_y$ and $\mean{B_y}$ are correlated. The slope of the
least-square fit gives a first estimate for $\alpha_{yy}$. It was found
that $\alpha_{yy}$ is negative in the upper disc plane and positive in
the lower, and its magnitude was about $0.001\Omega H$, where
$\Omega=2\pi/T_{\rm rot}$ is the local angular velocity and $H$ the
gaussian density scale height of the disc. In these papers no explicit
estimate for $\eta_{\rm t}$ was given, but one would expect that $\eta_{\rm t}$ is
comparable to the turbulent kinematic viscosity $\nu_{\rm t}$, which was found
to be approximately $0.005\Omega H^2$. The negative sign of $\alpha$ in
the upper disc plane was confirmed independently by Ziegler \& R\"udiger
(2000), who used a different code. Their simulations also produced a large
scale field. On the other hand, the simulations of Miller \& Stone (2000)
for a taller box did not show evidence for large scale dynamo action. It
is possible that in their case the effective magnetic Reynolds number
is larger and therefore the time scale after which a large scale field
can be established would be longer. This effect would be even more
pronounced if the numerical diffusion operator has hyper-diffusive
properties (Brandenburg \& Sarson 2002).

In writing down the mean field equations for the horizontally averaged
fields, we reiterate that because of periodic boundary conditions in the
horizontal directions the averaged vertical field $\mean{B_z}$, remains
unchanged, and since $\mean{B_z}=0$ initially, it remains so for all times.
Thus, we have only two equations for the mean fields in the $x$ and $y$
directions,
\EQ
{\partial\mean{B_x}\over\partial t}=-{\partial {\cal E}_y\over\partial z},
\label{dyneq1}
\EN
\EQ
{\partial\mean{B_y}\over\partial t}={\partial {\cal E}_x\over\partial z}
-q\Omega\mean{B_x}.
\label{dyneq1b}
\EN
The effect of molecular diffusion is here subsumed into the definition
of $\emf$. On the boundaries we assume
\EQ
{\partial\mean{B_x}\over\partial z}={\partial\mean{B_y}\over\partial z}=0
\quad\mbox{on}\quad z=0;\quad
\mean{B_x}=\mean{B_y}=0\quad\mbox{on}\quad z=L_z.
\label{dyneq1bc}
\EN
This is equivalent to the conditions used in the three-dimensional
simulations for the non-averaged magnetic field.  The last term in
\Eq{dyneq1b} arises from the local velocity shear,
$\mean{u_y}(x)=-q\Omega x$. In the present case of keplerian rotation
we have $q=3/2$; for estimates of $\alpha_{yy}$ for different values
of $q$ see Brandenburg \& Donner (1997). The simplest parameterization
that leads to dynamo action, balanced by diffusion, would be
\EQ
{\cal E}_x=\eta_{\rm t}{\partial\mean{B_y}\over\partial z},
\label{dyneq2b}
\EN
\EQ
{\cal E}_y=\alpha_{yy}\mean{B_y}-\eta_{\rm t}{\partial\mean{B_x}\over\partial z}.
\label{dyneq2}
\EN
It is instructive to consider first free wave solutions of the form
$e^{\lambda t+ikz}$, ignoring thus the boundaries. We now assume $k>0$.
The dispersion relation is
\EQ
\lambda=(1\pm i)|\textstyle{3\over4}\Omega\alpha_{yy}k|^{1/2}-\eta_{\rm t} k^2,
\label{dispersion}
\EN
where the two signs are respectively for positive and negative
values of $\alpha_{yy}$.
The marginally excited solution ($\mbox{Re}\lambda=0$) can be written as
\EQ
\mean{B_x}=A\sin k(z-ct),\quad
\mean{B_y}=\sqrt{2}\,A\left|{c\over\alpha_{yy}}\right|
\sin[k(z-ct)\pm{\textstyle{3\over4}}\pi],
\EN
where $A$ is the amplitude (undetermined in linear theory),
and $c$ is the wave speed,
\EQ
c\equiv-{{\rm Im}\lambda\over k}=
-\alpha_{yy}\left|{3\Omega\over4k\alpha_{yy}}\right|^{1/2}
=\mp\eta_{\rm t} k.
\EN
Note here that $c>0$ (as seen in the simulations, cf.\ \Fig{Fbutter})
requires $\alpha_{yy}<0$.
This is consistent with the sign of $\alpha_{yy}$ obtained earlier
by means of correlating ${\cal E}_y$ with $\mean{B_y}$
(BNST95, Brandenburg \& Donner 1997).

\epsfxsize=16cm\begin{figure}[t!]
\epsfbox{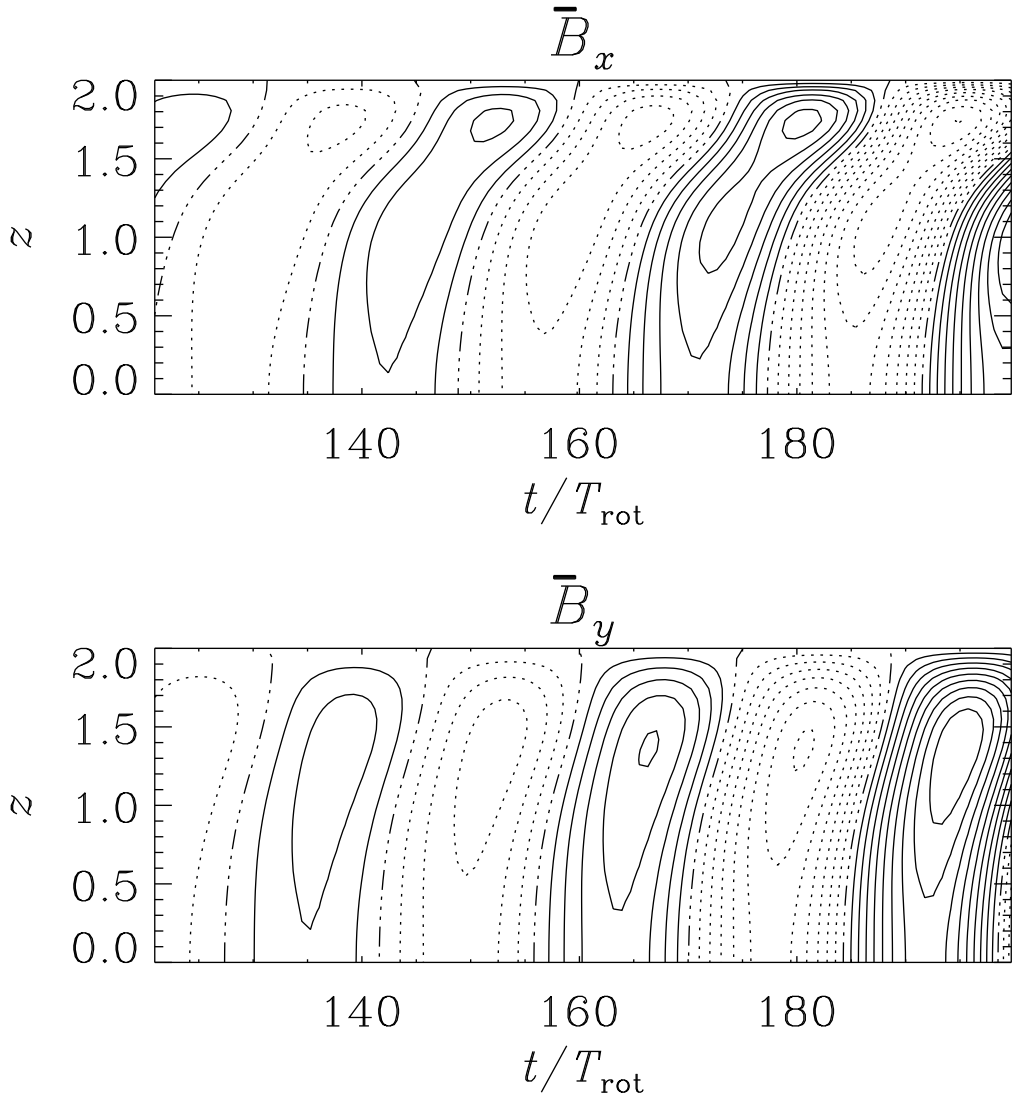}
\caption[]{Horizontally averaged radial ($\mean{B_x$}) and
toroidal ($\mean{B_y}$)
magnetic fields as a function of time (in rotational periods) and height, $z$,
as obtained from the one-dimensional dynamo model.
Dotted contours denote negative values.
}\label{Fmodel}\end{figure}

\epsfxsize=16cm\begin{figure}[t!]
\epsfbox{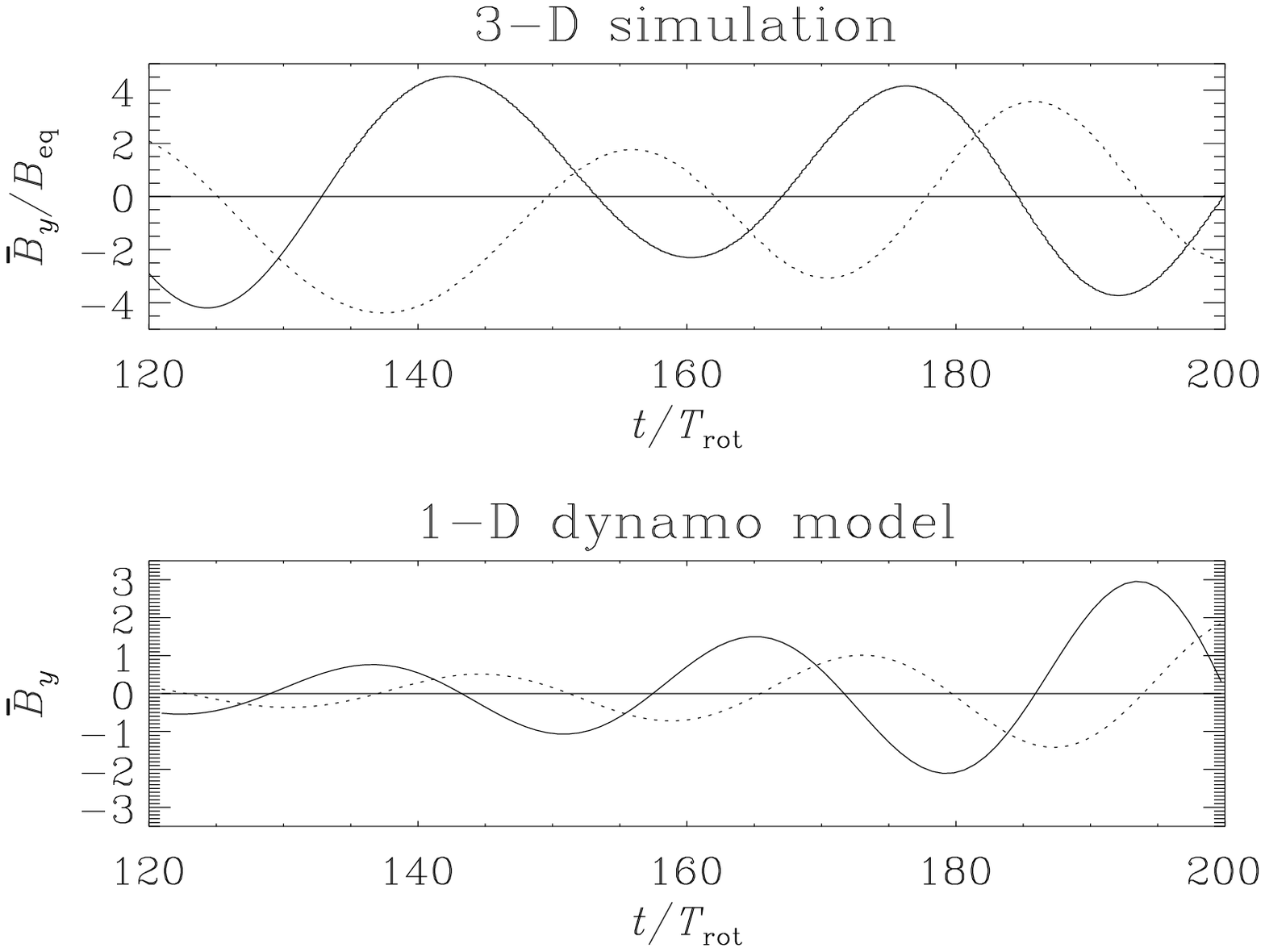}
\caption[]{
Comparison of $\mean{B_x}$ (dotted lines) and $\mean{B_y}$ (solid lines)
as functions of time for $z=1$ for the simulation and the model. Note the
similarity in the phase shifts between $\mean{B_x}$ and $\mean{B_y}$ in
the two cases. In both panels $\mean{B_x}$ has been scaled up by a factor of 20.
Note that both the model and the simulation have a similar amplitude
ratio between the two fields.
The horizontally averaged toroidal magnetic field in the simulation is given
in units of the equipartition value, $B_{\rm eq}=\bra{4\pi\rho\uu^2}^{1/2}$.
The model on the other hand is linear and the amplitude is therefore
undetermined. The field is growing in time, showing that the mean-field
dynamo model is supercritical.
}\label{Fphase}\end{figure}

In \Fig{Fmodel} we present the result of a numerical integration of
\eq{dyneq1}--\eq{dyneq2}, where the boundaries are now taken into account.
We chose
\EQ
\alpha_{yy}=-0.001\Omega z,\quad\eta_{\rm t}=0.005\Omega H^2,
\label{chosen}
\EN
with $L_z=2H$, and normalize to $\Omega=H=1$, so the rotational period
is then $T_{\rm rot}=2\pi$.

The model reproduces roughly the average behavior of the magnetic field
in the three-dimensional simulation.
Of course, the simulation shows strong fluctuations, which
are not explicitly present in mean field theory.
In the simulations fluctuations are natural because the flow is turbulent
and horizontal averages are only an approximation to ensemble averages.
Nevertheless the agreement between the mean-field model and the
turbulence simulation is striking and even quantitative: the cycle
period is $2\times15$ rotational periods, the migration speed is positive,
although about three times too fast ($c\approx+0.07\Omega H$), and the
phase difference between $\mean{B_x}$
and $\mean{B_y}$ is about $3\pi/4$ (Fig.~\ref{Fphase}).
Also the amplitude ratio between $\mean{B_x}$ and $\mean{B_y}$ is similar
for the simulation and the model (about 1/30). Thus, the only major
disagreement is in the migration speed, which is too fast in the model
compared to the simulation.

Our estimates of turbulent transport coefficients in \Eq{chosen} have
been purely phenomenological. What is missing is an analysis of the
{\it validity} of the expressions \eq{emf} or \eq{emf2}. This will be
attempted in the following section.

\section{Fitting the functional form of $\emf$}
\label{fitting}

In this section we attempt a direct determination of turbulent
transport coefficients. We now allow alpha and turbulent diffusivity to be tensors
whose components depend either on $z$ (local approach) or on the
vertical scale or wavenumber (nonlocal approach). In both approaches
we also consider the case where $\eta^*_{ij}$ is purely diagonal.

Here and elsewhere primes on $B_i$ and ${\cal E}_i$ denote
$z$-derivatives and dots denote time derivatives. We also drop the
overbars on $B_x$ and $B_y$, which denote now horizontally averaged
fields.

\subsection{A local formulation}
\label{Slocal}

The underlying turbulence is anisotropic and therefore alpha-effect
and turbulent diffusivity are really tensors. From \eq{emf} we have
\EQ
{\cal E}_x=\alpha_{xx}B_x+\alpha_{xy}B_y
+\eta_{xxz}B_x^\prime+\eta_{xyz}B_y^\prime,
\label{prodx}
\EN
\EQ
{\cal E}_y=\alpha_{yx}B_x+\alpha_{yy}B_y
+\eta_{yxz}B_x^\prime+\eta_{yyz}B_y^\prime.
\label{prody}
\EN
Since we know $\emf$ and $\BB$ at different times we can determine the
eight coefficients $\alpha_{ij}$ and $\eta_{ijz}$ by forming moments
with $B_x$, $B_y$, $B_x^\prime$, and $B_y^\prime$. This gives $2\times4$
equations for the 8 unknowns, $\alpha_{ij}$ and $\eta_{ijz}$.
The reason why the third index on $\eta_{ijk}$ is always $z$ is
because only the $z$-derivatives of $B_i$ are nonvanishing. This in
turn is because the horizontally averaged fields are independent of $x$
and $y$ (see previous section).

\begin{figure}[t!]
\epsfbox{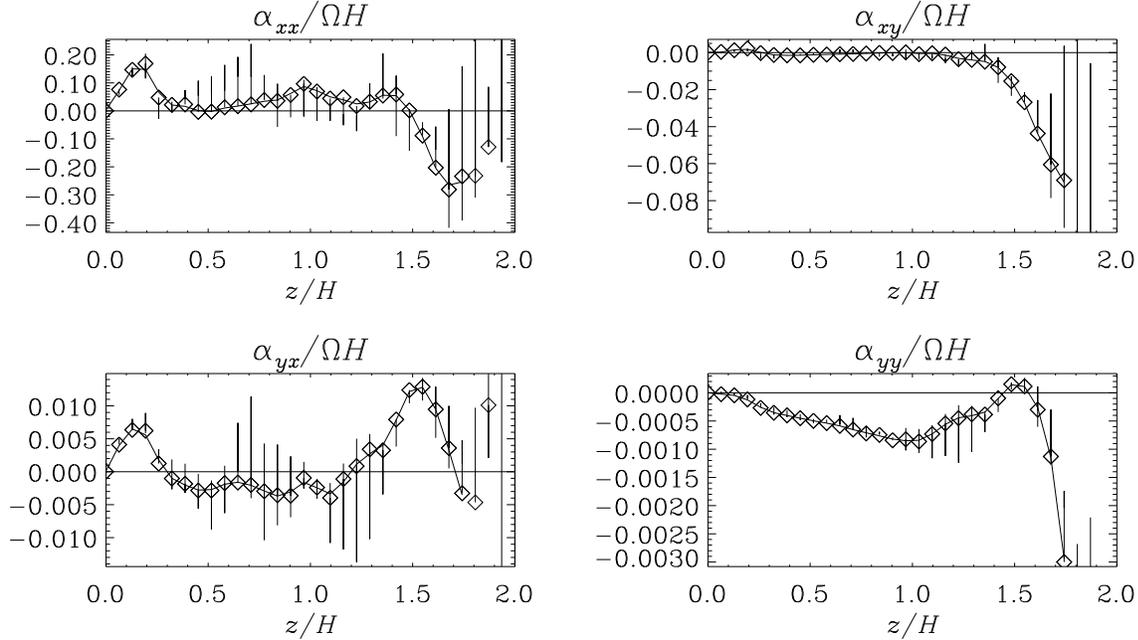}
\caption[]{The symbols denote the results for
$\alpha_{ij}(z)$ as obtained by solving \eq{matrixeqnz}
and the solid line represents a five point running mean.
}\label{Fpalp_loc}\end{figure}

The full system of equations can be written in the form of two
matrix equations,
\EQ
\EE^{(i)}(z)=\MM(z)\,\CC^{(i)}(z),\quad i=x,y,
\label{matrixeqnz}
\EN
with the matrix
\EQ
\MM=\pmatrix{
\bra{{B}_x{B}_x} &
\bra{{B}_x{B}_y} &
\bra{{B}_x{B}_x'} &
\bra{{B}_x{B}_y'} \cr
\bra{{B}_y{B}_x} &
\bra{{B}_y{B}_y} &
\bra{{B}_y{B}_x'} &
\bra{{B}_y{B}_y'} \cr
\bra{{B}_x'{B}_x} &
\bra{{B}_x'{B}_y} &
\bra{{B}_x'{B}_x'} &
\bra{{B}_x'{B}_y'} \cr
\bra{{B}_y'{B}_x} &
\bra{{B}_y'{B}_y} &
\bra{{B}_y'{B}_x'} &
\bra{{B}_y'{B}_y'}},
\EN
which is the same for both equations ($i=1$ and $i=2$), and the vectors
\EQ
\EE^{(i)}=\pmatrix{
\bra{{\cal E}_i{B}_x}\cr
\bra{{\cal E}_i{B}_y}\cr
\bra{{\cal E}_i{B}_x'}\cr
\bra{{\cal E}_i{B}_y'}},\quad
\CC^{(i)}=\pmatrix{
{\alpha_{ix}}\cr
{\alpha_{iy}}\cr
{\eta_{ixz}}\cr
{\eta_{iyz}}}.
\label{resultz}
\EN
The averages are taken over time. In \Figs{Fpalp_loc}{Fpeta_loc} we
show the results respectively for the coefficients $\alpha_{ij}$ and
$\eta_{ijz}$ as functions of $z$. We also plot as a solid line a five
point running mean of the data.

\epsfxsize=16cm

\epsfxsize=16cm\begin{figure}[t!]
\epsfbox{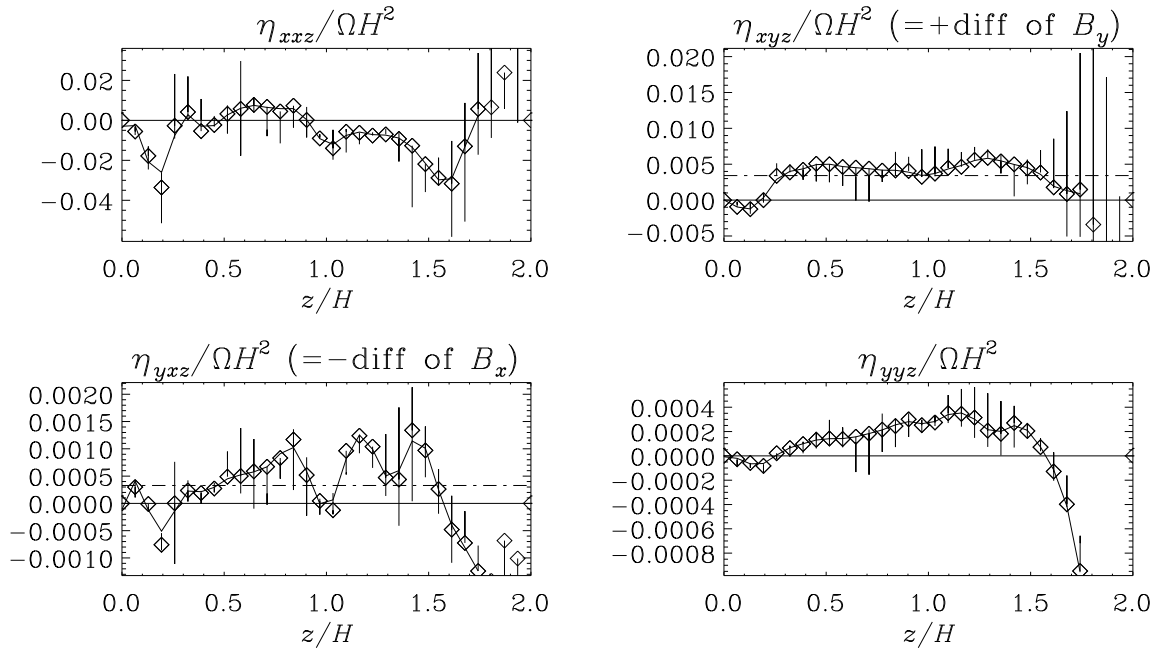}
\caption[]{The symbols denote the results for
$\eta_{ijz}(z)$ as obtained by solving \eq{matrixeqnz}
and the solid line represents a five point running mean.
The two horizontal dash-dotted lines give the average values.
}\label{Fpeta_loc}\end{figure}

Here and below our error bars are a measure of the stability of the
fit when only part of the time series is used. In practice we divide
the data set into two parts and calculate the transport coefficients
separately for each subset. The error bars cover then the range of
values obtained by using all or only part of the data. The error bars
would probably shrink if we extended the data set, but some of the
noise is due to fluctuations from cycle to cycle and therefore
physical.

Near the top boundary the data points
deviate significantly from the relatively smooth trend seen in the
data away from the boundary. Therefore we consider the data near the
boundary as uncertain and, when calculating averages and running means,
we ignore all points within four mesh zones near the boundary. The
vertically averaged values of the components of $\alpha_{ij}$ and
$\eta_{ijz}$ are
\EQ
\alpha_{ij}=\pmatrix{0.010 &-0.009\cr 0.001 & -0.001}\Omega H,\quad
\eta_{ijz}=\pmatrix{-0.007 & 0.003\cr 0.000 & 0.000}\Omega H^2.
\EN
Note that $\alpha_{yy}=-0.001\Omega H$ and $\eta_{xyz}=0.003\Omega
H^2$, in rough agreement with the results found earlier from other
considerations.  However, $-\eta_{yxz}$ is at least ten times smaller and
perhaps even negative. This coefficient is responsible for the diffusion
of $B_x$, i.e.\ $\dot{B}_x=...-\eta_{yxz}B_x^{\prime\prime}$. If
$-\eta_{yxz}<0$, this would indicate that some higher order terms
(hopefully with the right sign!) would need to be restored for
stabilization. Typically, an infinite series of further terms could
then become important (Dittrich et al.\ 1984, Elperin et al.\ 2000),
making this whole approach difficult to use in practice. We address this
difficulty in \Secs{Snonlocal}{Snonlocalstar} by adopting a nonlocal
approach.

\subsection{A local formulation using a diagonal diffusion tensor}
\label{Slocaldiag}

\begin{figure}[t!]
\epsfbox{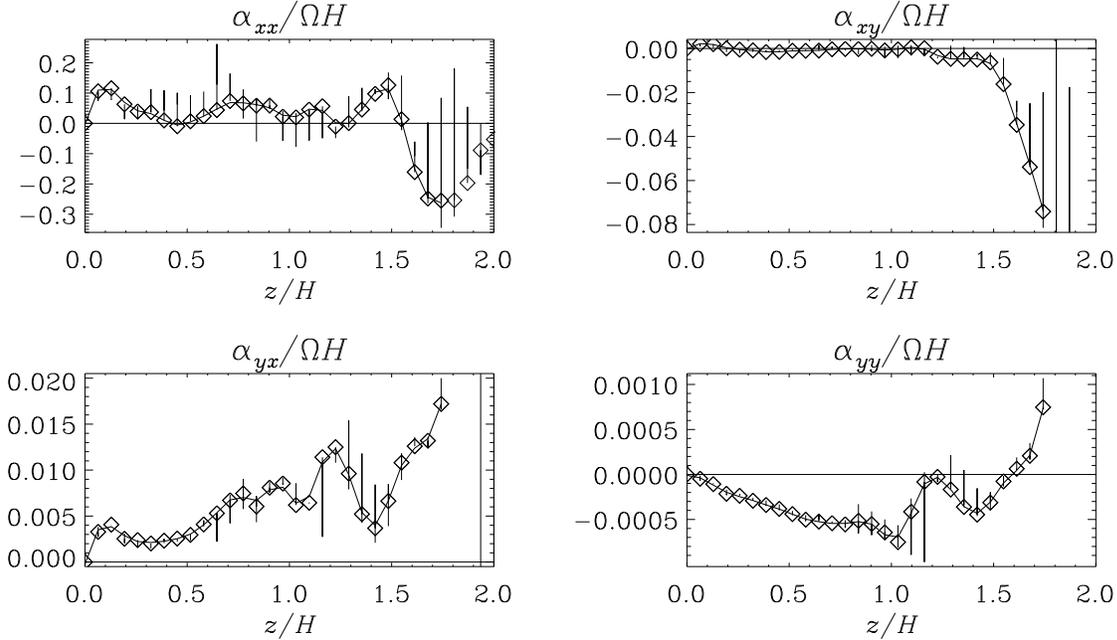}
\caption[]{The symbols denote the results for
$\alpha_{ij}(z)$ as obtained by solving \eq{matrixeqnzJ}
and the solid line represents a five point running mean.
Here and below the error bars indicate the stability of the
fit when only a subset of the data is used.
}\label{Fpalp_locJ}\end{figure}

We now adopt the $J$-formulation for the diffusion term,
i.e.\ \Eq{emf2}, using a diagonal diffusion tensor,
i.e.\ $\eta^*_{xy}=\eta^*_{yx}=0$.  In that case we have to solve two
systems of {\it three} (instead of four) equations,
\EQ
\EE^{(i)}(z)=\MM^{(i)}(z)\,\CC^{(i)}(z),\quad i=x,y,
\label{matrixeqnzJ}
\EN
where
\EQ
\MM^{(i)}=\pmatrix{
\bra{{B}_x{B}_x} &
\bra{{B}_x{B}_y} &
-\bra{{B}_x{J}_i} \cr
\bra{{B}_y{B}_x} &
\bra{{B}_y{B}_y} &
-\bra{{B}_y{J}_i} \cr
-\bra{{J}_i{B}_x} &
-\bra{{J}_i{B}_y} &
\bra{{J}_i{J}_i}}
\EN
is now different for the two equations ($i=1$ and $i=2$).
(Here and below, {\it no} summation over $i$ is implied!) The vectors
$\EE^{(i)}$ are given by
\EQ
\EE^{(i)}=\pmatrix{
\bra{{\cal E}_i{B}_x}\cr
\bra{{\cal E}_i{B}_y}\cr
-\bra{{\cal E}_i{J}_i}},\quad
\CC^{(i)}=\pmatrix{
{\alpha_{ix}}\cr
{\alpha_{iy}}\cr
{\eta^*_{ii}}}.
\label{resultzJ}
\EN
Again, the averages are taken
over time. In \Figs{Fpalp_locJ}{Fpeta_locJ} we show the results
respectively for the coefficients $\alpha_{ij}$ and $\eta^*_{ij}$ as
functions of $z$. We also plot as a solid line a five point running mean
of the data.

\epsfxsize=16cm

\epsfxsize=16cm\begin{figure}[t!]
\epsfbox{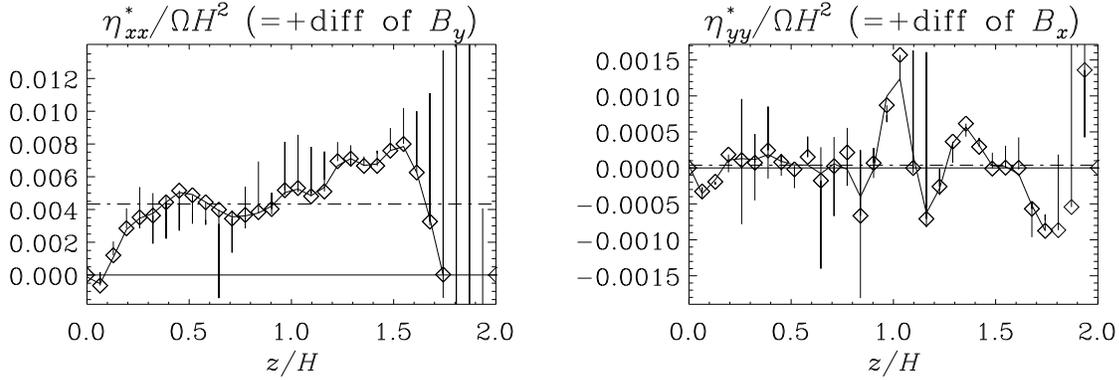}
\caption[]{The symbols denote the results for
$\eta_{ij}^*(z)$ as obtained by solving \eq{matrixeqnzJ}
and the solid line represents a five point running mean.
The average values are indicated by horizontal dash-dotted lines.
}\label{Fpeta_locJ}\end{figure}

Note that both $\eta^*_{xx}$ and $\eta^*_{yy}$ are now positive on
average (indicated by the dash-dotted line). Nevertheless, $\eta^*_{yy}$,
which is responsible for the diffusion of $B_x$, shows still large
fluctuations and can locally still be negative. Taken at face value,
such excursions of $\eta^*_{yy}$ to negative values would cause a
catastrophic growth of small-scale structures in a mean-field model.
We note that the relation \eq{relation} between the different components
of $\eta^*_{ij}$ and $\eta_{ijz}$ still holds approximately, even though
this relation is no longer strictly valid if, as in the present case,
the fit to the diagonal components of $\eta^*_{ij}$ is obtained under
the restriction that the off-diagonal components vanish.

\subsection{A nonlocal formulation}
\label{Snonlocal}

We recall that \Eq{emf} corresponds really to a Taylor expansion of
an underlying integral kernel in terms of derivatives of delta functions
(see, e.g., Hasler, Kaisig \& R\"udiger 1995; see also Nicklaus \& Stix 1988).
Therefore we now write equations \eq{prodx} and \eq{prody} as a convolution
in the form
\EQ
{\cal E}_x=\alpha_{xx}* B_x+\alpha_{xy}* B_y
+\eta_{xxz}* B_x^\prime+\eta_{xyz}* B_y^\prime,
\label{convx}
\EN
\EQ
{\cal E}_y=\alpha_{yx}* B_x+\alpha_{yy}* B_y
+\eta_{yxz}* B_x^\prime+\eta_{yyz}* B_y^\prime,
\label{convy}
\EN
where $\alpha_{ij}(z,z',t)$ and $\eta_{ijz}(z,z',t)$ are now integral kernels
and the asterisks refer to a convolution,
\EQ
\alpha_{ij}* B_j\equiv\int_0^{L_z}\alpha_{ij}(z,z')B_j(z',t)dz',
\label{conva}
\EN
\EQ
\eta_{ijz}* B_j^\prime\equiv\int_0^{L_z}\eta_{ijz}(z,z')B_j^\prime(z',t)dz'.
\label{conve}
\EN
We solve for the 8 kernels using Fourier transformation. We introduce
\EQ
\hat{\cal E}_i'=\int_0^{L_z}{\cal E}_i'(z)\cos kz\,dz,
\EN
\EQ
\hat{B}_i=\int_0^{L_z}B_i(z)\cos kz\,dz,
\EN
where
\EQ
k=k_n=(n+1/2)\pi/L_z,\quad n=0,1,...
\EN
Note that we use here the Fourier transform of the {\it derivative}
of $\emf$, which is more easily obtained from the simulations.
We assume that $\alpha_{ij}(z,z')$ and $\eta_{ijz}(z,z')$ have the form
\EQ
\alpha_{ij}(z,z')=\sin k_0z\,(2/L_z)\sum_{k}\sin kz\sin kz'\,\hat\alpha_{ij}(k),
\label{alpk}
\EN
\EQ
\eta_{ijz}(z,z')=(2/L_z)\sum_{k}\sin kz\sin kz'\,\hat\eta_{ijz}(k),
\label{etak}
\EN
where $k_0=\pi/(2L_z)$. We also need
\EQ
\tilde{B}_i(k)=\int_0^{L_z}B_i(z)\sin k_0z\,\sin kz\,dz,
\EN
and the corresponding inverse transform in terms of $\tilde{B}_i(k)$,
\EQ
B_i(z)=(2/L_z)(\sin k_0 z)^{-1}\sum_k\sin kz\tilde{B}_i(k).
\EN
The $\sin k_0z$ term in \eq{alpk} ensures that the $\alpha$-effect
is antisymmetric about the equatorial plane. For example, if
$\hat\alpha_{ij}(k)$ is independent of $k$, then the integral in
\Eq{conva} simply becomes a multiplication, so
$\alpha_{ij}* B_j=\sin k_0z\,\alpha_{ij}^{(0)}B_j(z)$,
where $\alpha_{ij}^{(0)}=\hat\alpha_{ij}(k)$ is a constant,
independent of $k$.
We are now able to write \eq{convx} and \eq{convy} in the following form
\EQ
\hat{\cal E}_x'=k\hat\alpha_{xx}\tilde{B}_x+k\hat\alpha_{xy}\tilde{B}_y
-k^2\hat\eta_{xxz}\hat{B}_x-k^2\hat\eta_{xyz}\hat{B}_y,
\label{fourx}
\EN
\EQ
\hat{\cal E}_y'=k\hat\alpha_{yx}\tilde{B}_x+k\hat\alpha_{yy}\tilde{B}_y
-k^2\hat\eta_{yxz}\hat{B}_x-k^2\hat\eta_{yyz}\hat{B}_y.
\label{foury}
\EN

\begin{figure}[t!]
\epsfbox{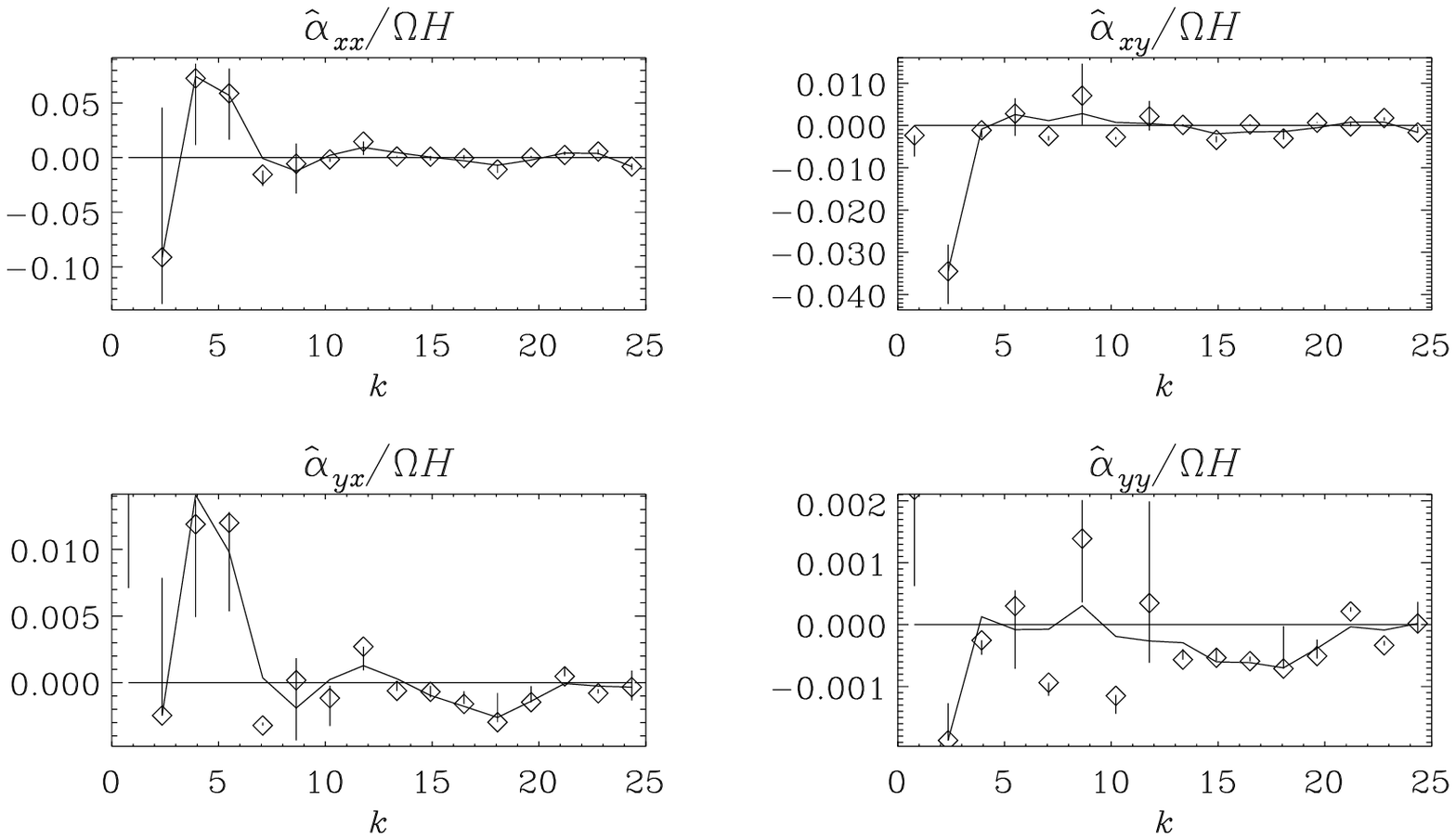}
\caption[]{The symbols denote the results for
$\hat\alpha_{ij}(k)$ as obtained from \eq{matrixeqnk}
and the solid line represents a five point running mean
ignoring the point $k=k_0$.
}\label{Fpalp}\end{figure}

Again, we can solve for the 8 coefficients $\hat\alpha_{ij}$ and
$\hat\eta_{ijz}$ by computing moments (separately for each value of $k$).
We obtain two matrix equations of the form
\EQ
\tilde{\EE}^{(i)}(k)=\tilde{\MM}(k)\,\tilde{\CC}^{(i)}(k),\quad i=x,y
\label{matrixeqnk}
\EN
with the matrix
\EQ
\tilde{\MM}=\pmatrix{
\bra{\tilde{B}_x\tilde{B}_x} &
\bra{\tilde{B}_x\tilde{B}_y} &
\bra{\tilde{B}_x\hat{B}_x} &
\bra{\tilde{B}_x\hat{B}_y} \cr
\bra{\tilde{B}_y\tilde{B}_x} &
\bra{\tilde{B}_y\tilde{B}_y} &
\bra{\tilde{B}_y\hat{B}_x} &
\bra{\tilde{B}_y\hat{B}_y} \cr
\bra{\hat{B}_x\tilde{B}_x} &
\bra{\hat{B}_x\tilde{B}_y} &
\bra{\hat{B}_x\hat{B}_x} &
\bra{\hat{B}_x\hat{B}_y} \cr
\bra{\hat{B}_y\tilde{B}_x} &
\bra{\hat{B}_y\tilde{B}_y} &
\bra{\hat{B}_y\hat{B}_x} &
\bra{\hat{B}_y\hat{B}_y}},
\EN
which is the same for both equations ($i=1$ and $i=2$), and the vectors
\EQ
\tilde{\EE}^{(i)}=\pmatrix{
\bra{\hat{\cal E}_i'\tilde{B}_x}\cr
\bra{\hat{\cal E}_i'\tilde{B}_y}\cr
\bra{\hat{\cal E}_i'\hat{B}_x}\cr
\bra{\hat{\cal E}_i'\hat{B}_y}},\quad
\tilde{\CC}^{(i)}=\pmatrix{
{k\hat\alpha_{ix}}\cr
{k\hat\alpha_{iy}}\cr
{-k^2\hat\eta_{ixz}}\cr
{-k^2\hat\eta_{iyz}}}.
\label{result}
\EN
The averages are taken over time.
In \Figs{Fpalp}{Fpeta} we show the results respectively for the coefficients
$\hat\alpha_{ij}$ and $\hat\eta_{ijz}$ as functions of $k$. We also plot
as a solid line a five point running mean of the data. On the boundaries
the lines go through the original data points, however.

\epsfxsize=16cm

\epsfxsize=16cm\begin{figure}[t!]
\epsfbox{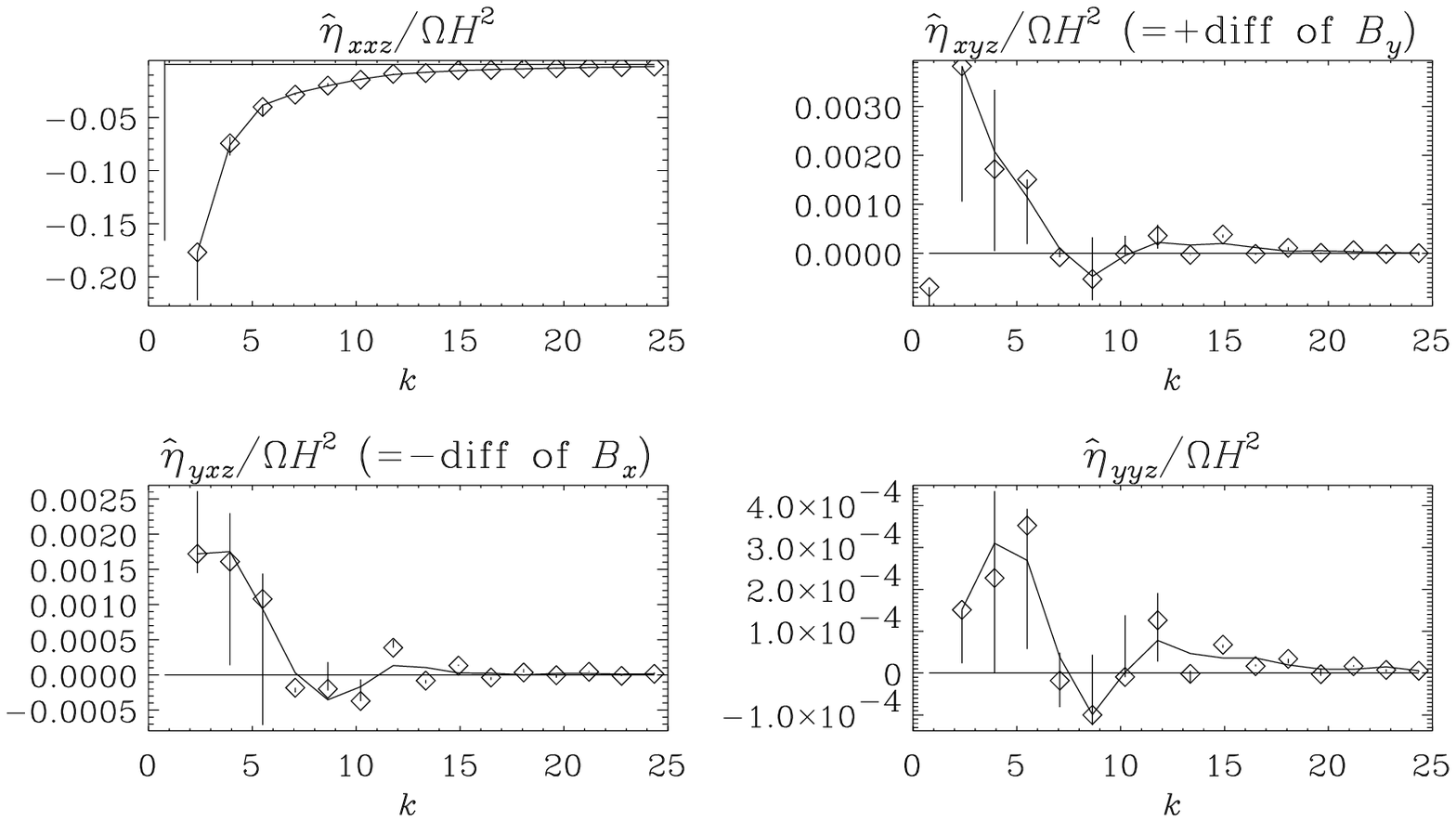}
\caption[]{The symbols denote the results for
$\hat\eta_{ijz}(k)$ as obtained from \eq{matrixeqnk}
and the solid line represents a five point running mean.
Note that the data points for the smallest value of $k$,
$k=k_0$, deviate strongly from those for $k_1$ and lie
outside the plot ($\hat\eta_{xxz}(k_0)=0.13\Omega H^2$,
$\hat\eta_{xyz}(k_0)=-0.0007\Omega H^2$,
$\hat\eta_{yxz}(k_0)=0.04\Omega H^2$,
and $\hat\eta_{yy}^*(k_0)=0.002\Omega H^2$).
}\label{Fpeta}\end{figure}

In the traditional $\alpha\Omega$-dynamo the $(y,y)$-component is the
most important one. However,
$\hat\alpha_{yy}(k)$ shows substantial variations in $k$, indicating
possibly large uncertainties, although $\hat\alpha_{yy}$ is mostly
negative and the value for $k_1$ is around $-0.001$, similar to the
value we used in \Eq{chosen}. For the diffusion coefficient
$\hat\eta_{xyz}$ the value for $k=k_1$ is positive, i.e.\ stabilizing.
Furthermore, the value for $k=k_1$ is much larger than for the next
higher values of $k$. One possible suggestion would be to assume that
$\eta_{ijz}(k)$ was proportional to $k^{-2}$. However, this would
correspond to diffusion of the algebraic form
$\dot{B}_i=...-B_i/\tau^{(i)}$, i.e.\ without any differential
(or integral) operator. Here the superscript $(i)$ on $\tau$ indicates
that the decay times are in general different for the $x$ and $y$
directions. Let us recall, however, that because the eigenvalues of the
diffusion tensor are not all positive our fitting is still not self-consistent.

\subsection{A nonlocal formulation using a diagonal diffusion tensor}
\label{Snonlocalstar}

\begin{figure}[t!]
\epsfbox{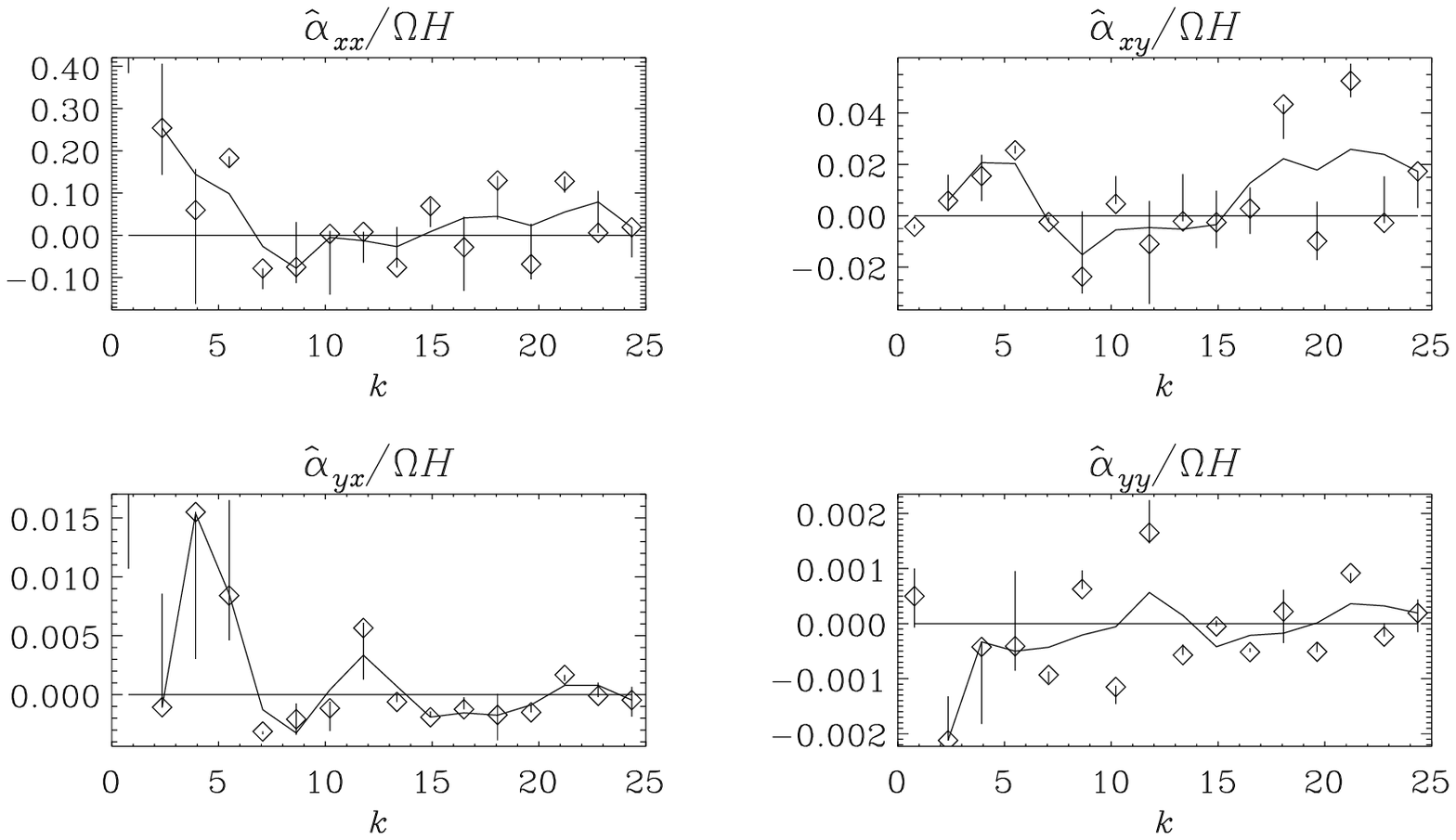}
\caption[]{The symbols denote the results for
$\hat\alpha_{ij}(k)$ as obtained from \eq{matrixeqnkJ}
and the solid line represents a five point running mean
ignoring the point $k=k_0$.
}\label{FpalpJ}\end{figure}

Like in \Sec{Slocaldiag} we now adopt the formulation \eq{emf2} using
a diagonal diffusion tensor. We write
\EQ
\hat{\cal E}_x'=k\hat\alpha_{xx}\tilde{B}_x+k\hat\alpha_{xy}\tilde{B}_y
-\hat\eta^*_{xx}\hat{J}_x',
\label{fourxJ}
\EN
\EQ
\hat{\cal E}_y'=k\hat\alpha_{yx}\tilde{B}_x+k\hat\alpha_{yy}\tilde{B}_y
-\hat\eta^*_{yy}\hat{J}_y',
\label{fouryJ}
\EN
where $\hat{J}_x'=+k^2\hat{B}_y$ and $\hat{J}_y'=-k^2\hat{B}_x$.
Instead of \Eq{matrixeqnk} we have
\EQ
\tilde{\EE}^{(i)}(k)=\tilde{\MM}^{(i)}\!(k)\,\tilde{\CC}^{(i)}(k),\quad i=x,y
\label{matrixeqnkJ}
\EN
with the matrix
\EQ
\tilde{\MM}^{(i)}=\pmatrix{
\bra{\tilde{B}_x\tilde{B}_x} &
\bra{\tilde{B}_x\tilde{B}_y} &
-\bra{\tilde{B}_x\hat{J}_i'} \cr
\bra{\tilde{B}_y\tilde{B}_x} &
\bra{\tilde{B}_y\tilde{B}_y} &
-\bra{\tilde{B}_y\hat{J}_i'} \cr
-\bra{\hat{J}_i'\tilde{B}_x} &
-\bra{\hat{J}_i'\tilde{B}_y} &
\bra{\hat{J}_i'\hat{J}_i'}}.
\EN
(Again, here and below no summation over $i$ is implied!)
The vectors are
\EQ
\tilde{\EE}^{(i)}=\pmatrix{
\bra{\hat{\cal E}_i'\tilde{B}_x}\cr
\bra{\hat{\cal E}_i'\tilde{B}_y}\cr
-\bra{\hat{\cal E}_i'\hat{J}_i'}},\quad
\tilde{\CC}^{(i)}=\pmatrix{
{k\hat\alpha_{ix}}\cr
{k\hat\alpha_{iy}}\cr
{\hat\eta^*_{ii}}}.
\label{resultJ}
\EN
In \Figs{FpalpJ}{FpetaJ} we show the results respectively for the coefficients
$\hat\alpha_{ij}$ and $\hat\eta^*_{ij}$ as functions of $k$. We also plot
as a solid line a five point running mean of the data.

\epsfxsize=16cm

\epsfxsize=16cm\begin{figure}[t!]
\epsfbox{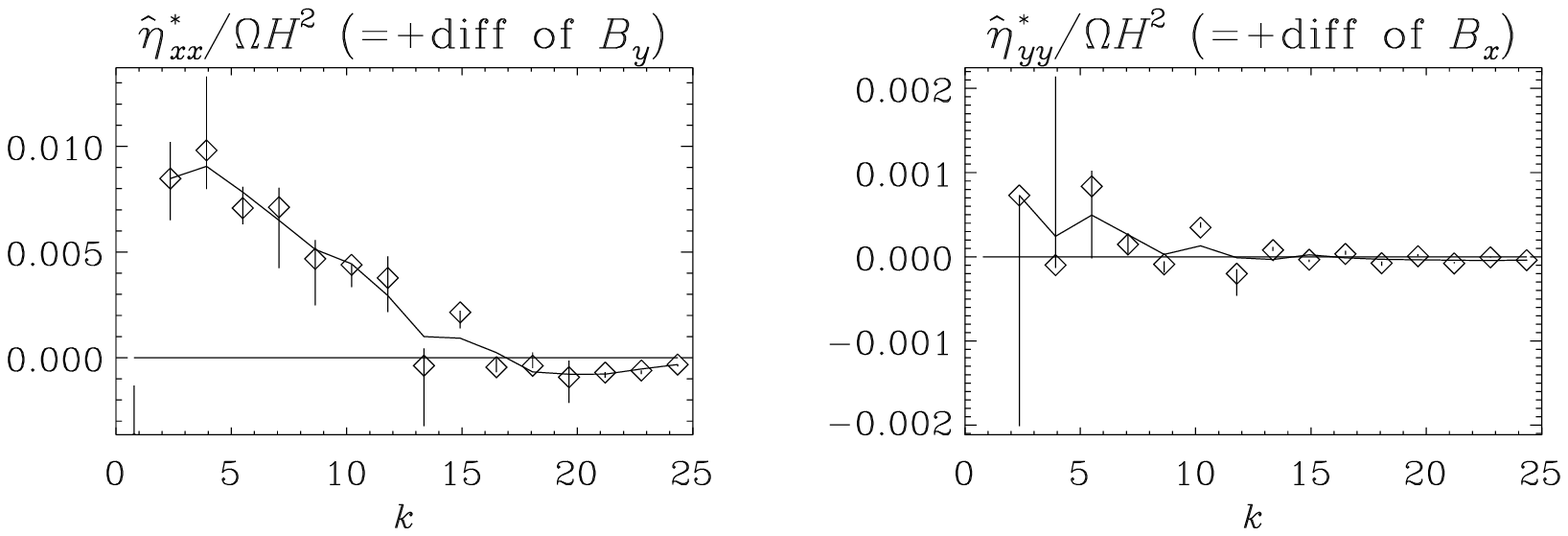}
\caption[]{The symbols denote the results for
$\hat\eta_{ij}^*(k)$ as obtained from \eq{matrixeqnkJ}
and the solid line represents a five point running mean.
Note that the data points for the smallest value of $k$,
$k=k_0$, deviate strongly from those for $k_1$ and lie
outside the plot ($\hat\eta_{xx}^*(k_0)=-0.004\Omega H^2$
and $\hat\eta_{yy}^*(k_0)=0.01\Omega H^2$).
}\label{FpetaJ}\end{figure}

Now both $\hat\eta^*_{xx}(k)$ and $\hat\eta^*_{yy}(k)$ are positive
in the range $1\la k\la 15$. Like in \Sec{Slocaldiag} the restriction
to a purely diagonal magnetic diffusion tensor has helped to make
turbulent diffusion positive definite. For larger values of $k$,
corresponding to smaller scales, the data are probably no longer
reliable. For the smallest value of $k$, $k=k_0$, the diffusion
coefficient of $B_y$ is negative, corresponding to a destabilization
of the field on large scales. While this is of course what is seen
in the simulation, it is unclear whether this is really due to a
negative diffusion on large scales (like in the Kuramoto-Sivashinsky
equation with negative diffusion and positive hyperdiffusion), or
just a natural manifestation of the $\alpha$-effect or an inverse cascade.
Let us note that the negative diffusion in the largest scale does not
lead to any catastrophic behavior of the solution. For recent work on
the negative diffusion effect see Zheligovsky \ea (2001).

\subsection{Comparative remarks}

In the previous subsections we have seen that $\alpha$-effect and
turbulent diffusion are not only always tensors, but their components
show also systematic spatial variations and, more importantly, they
are dominated by only the smallest wavenumbers. It does not seem
feasible, however, to determine simultaneously spatial variations and
wavenumber dependence. We emphasize that we do not advocate that the
model of \Sec{Snonlocalstar} is better than that of \Sec{Slocaldiag}
or even \Sec{Scomparison},
for example. Instead, elements of both approaches should preferentially
be taken into consideration.

Generally, we do find, however, that the restriction to only diagonal
components of the $\eta_{ij}^*(z)$ or $\hat\eta_{ij}^*(k)$ tensors is
to be preferred. This is because, on physical grounds, these diagonal
components should be positive. If one does allow for off-diagonal
components, as in \Secs{Slocal}{Snonlocal}, one finds (as expected)
somewhat different
results ($-\eta_{yxz}$ is different from $\eta_{yy}^*$, for example)
and diffusion of $B_x$ field is then mostly negative.

We wish to emphasize the importance of simultaneously determining
the various transport coefficients, as we have attempted here. To our
knowledge, the only other time this was done was in B01 where
scalar $\alpha$ and $\eta_{\rm t}$ quenchings have been determined.
In B01, both direct determination and fitting to a model were used
to obtain transport coefficients. Of course, fitting to a model is a
much more stable and reliable procedure, but it is necessarily model
dependent and one still has to assess whether or not the model is actually
consistent with the data.

An important complication arises from the fact that the actual
electromotive force contains a strongly fluctuating component which is
physical. This affects the results obtained from a direct determination
of transport coefficients and, consequently, many of the detailed
features cannot be physically meaningful. To our knowledge, dynamo
models with nonlocal (wavenumber dependent) transport coefficients were
never considered before. In order to have a preliminary assessment of
such models, and to obtain insight as to which components are crucial,
we now consider simple model calculations with wavenumber dependent
transport coefficients.

\section{Model calculations}

\epsfxsize=16cm\begin{figure}[t!]
\epsfbox{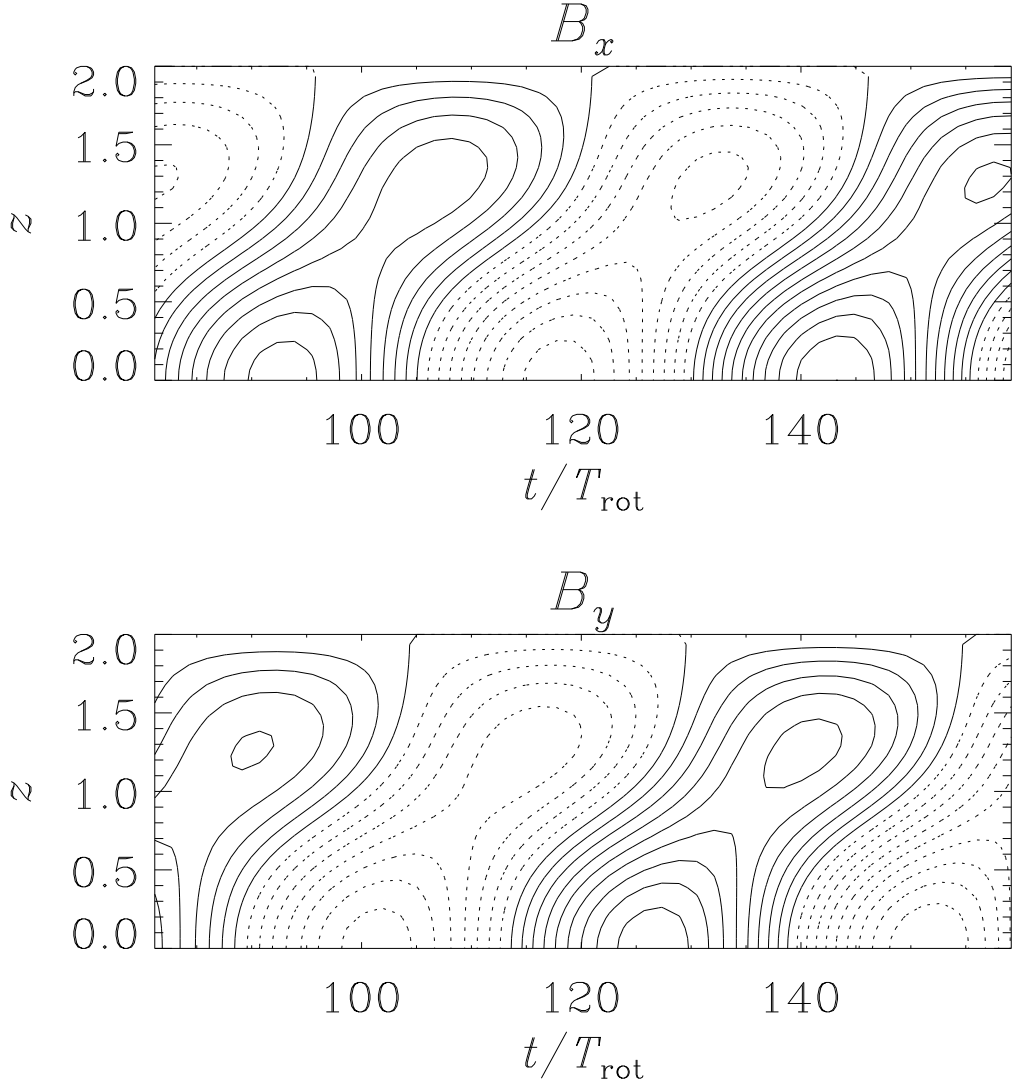}
\caption[]{
Model with $\hat\alpha_{yy}(k)=-0.001(1-k/k_{\rm max})$,
$k\leq k_{\rm max}=4$, $\tau^{-1}=0.01$.
}\label{Fnonloc}\end{figure}

The direct determination of nonlocal transport coefficients used in
\Sec{fitting} suggests that nonlocal effects could be important.
In fact, the magnitudes of $\hat\alpha_{ij}$ and $\hat\eta^*_{ij}$
tend to decrease with $k$. A $k^{-2}$ dependence of $\hat\eta^*_{xx}$
and $\hat\eta^*_{yy}$ would cancel the $\nabla^2$ diffusion operator
and would yield a simple (and again local!) $-1/\tau$ damping term.
The problem with this type of damping is clear when looking at the
dispersion relation \eq{dispersion}. For simplicity we assume a scalar
magnetic diffusivity,
$\hat\eta_{xx}^*(k)=\hat\eta_{yy}^*(k)\equiv\hat\eta_{\rm t}(k)$,
and that $\hat\eta_{\rm t}(k)=(\tau k^2)^{-1}$ we have
\EQ
\lambda(k)=(1\pm i)
|\textstyle{3\over4}\Omega\hat\alpha_{yy}(k)k|^{1/2}-\tau^{-1},
\label{dispersion2}
\EN
so for $\hat\alpha_{yy}(k)=\mbox{const}$, we would have $\lambda(k)>0$
for some $k>k_{\rm crit}$.
In order to prevent such an `ultraviolet catastrophe' we have to require
that $\hat\alpha_{yy}(k)\rightarrow0$ for large values of $k$. On the other
hand, the product $\hat\alpha_{yy}(k)k$ should reach its maximum for
relatively small values of $k$, because the simulations (Fig.~\ref{Fbutter})
suggested no systematic high-$k$ dependence.

Looking at \Fig{FpalpJ}, $\hat\alpha_{yy}(k)$ fluctuates essentially
about zero, although it does show a negative peak at $k=3\pi/4\approx2.4$.
Although $|\hat\alpha_{xx}|$ is much larger than $|\hat\alpha_{yy}|$,
it is relatively unimportant because it multiplies $B_x$, which is small,
and $\hat\alpha_{xx}$ only contributes to generating $B_y$ field, which
is more effectively generated by the shear term. Given that it is only
$\hat\alpha_{yy}(k)$ which contributes to $\alpha\Omega$-type dynamo
action, we need to model this term somehow. We do this by assuming a
linear dependence on $k$ for $k\le4$ of the form
\EQ
\hat\alpha_{yy}(k)=-0.001(1-k/k_{\rm max})\quad\mbox{for}\quad
k\leq k_{\rm max}=4.
\label{approx_alp}
\EN
For $\hat\eta_{\rm t}(k)$ we assume
\EQ
\hat\eta_{\rm t}(k)=(\tau k^2)^{-1}\quad\mbox{with}\quad
\tau^{-1}=0.01.
\label{approx_eta}
\EN
The resulting butterfly diagram is quite reasonable;
see Fig.~\ref{Fnonloc}. The main difference
compared with the local model shown in \Fig{Fmodel} is that the dynamo
wave migrates now slower. The fast migration was the main shortcoming
of the local model, where field reversals happened almost
simultaneously both at the midplane and at the upper boundary.

The assumption of a scalar $\hat\eta_{\rm t}(k)$ is made purely
for simplicity. For comparison we have also considered a model
with $\hat\eta_{yy}^*(k)=0.1\hat\eta_{xx}^*(k)$ (as suggested
by \Fig{FpetaJ}) and $\hat\eta_{xx}^*(k)=(\tau k^2)^{-1}$ with
$\tau^{-1}=0.01$ (as before). In order that the model is still
only marginally excited, we had to reduce $\hat\alpha_{yy}(k)$
by a uniform scaling factor of 5. Except for an increase of the
cycle period by about a factor of 2, the resulting butterfly diagram
was almost the same as in (Fig.~\ref{Fnonloc}).

Given that (i) this approximation reproduces phenomenology,
(ii) is based on a fit to the actual electromotive force (see
\Figs{FpalpJ}{FpetaJ}), and (iii) the dissipation is positive definite,
we consider \Eqs{approx_alp}{approx_eta} as the most reasonable
parameterization if one wanted to use mean-field theory.

The results for many of the components of the $\alpha$- and $\eta$-tensors
are rather noisy. However, the fact that the obtained values for
$\alpha_{yy}$, $\hat\alpha_{yy}(k)$, and $\eta_{\rm t}$ and $(\tau
k^2)^{-1}$, are all of the same sign and of comparable magnitude suggests
this is a stable result of the analysis. This result is also in agreement
with what is required for a mean-field model to work. Firstly,
the sign of $\hat\alpha_{yy}$ has to be negative
to make the dynamo wave move with a positive migration speed.
Secondly, the magnitude of $\hat\alpha_{yy}$ has to be around $10^{-3}$
so that the dynamo period is around 30 rotational periods. Finally, $\eta_{\rm t}$
has to be around $10^{-2}$ so that the dynamo is just weakly supercritical.

\section{Noise in the transport coefficients}

The plots for both local and nonlocal approaches show strong fluctuations
in the various transport coefficients. This is caused by strong noise, which
is much more pronounced in the electromotive force than in the resulting
mean field. It would therefore be surprising if the $\alpha$-effect was
really responsible for generating the much less noisy large scale magnetic
field in the simulation. However, it is important to note that even a
completely noisy $\alpha$-effect with no net component can amplify magnetic
fields (Moffatt 1978). Furthermore, in the presence of large scale shear
it is possible to generate also a large scale field (Vishniac \& Brandenburg
1997). 

\begin{figure}[t!]
\epsfbox{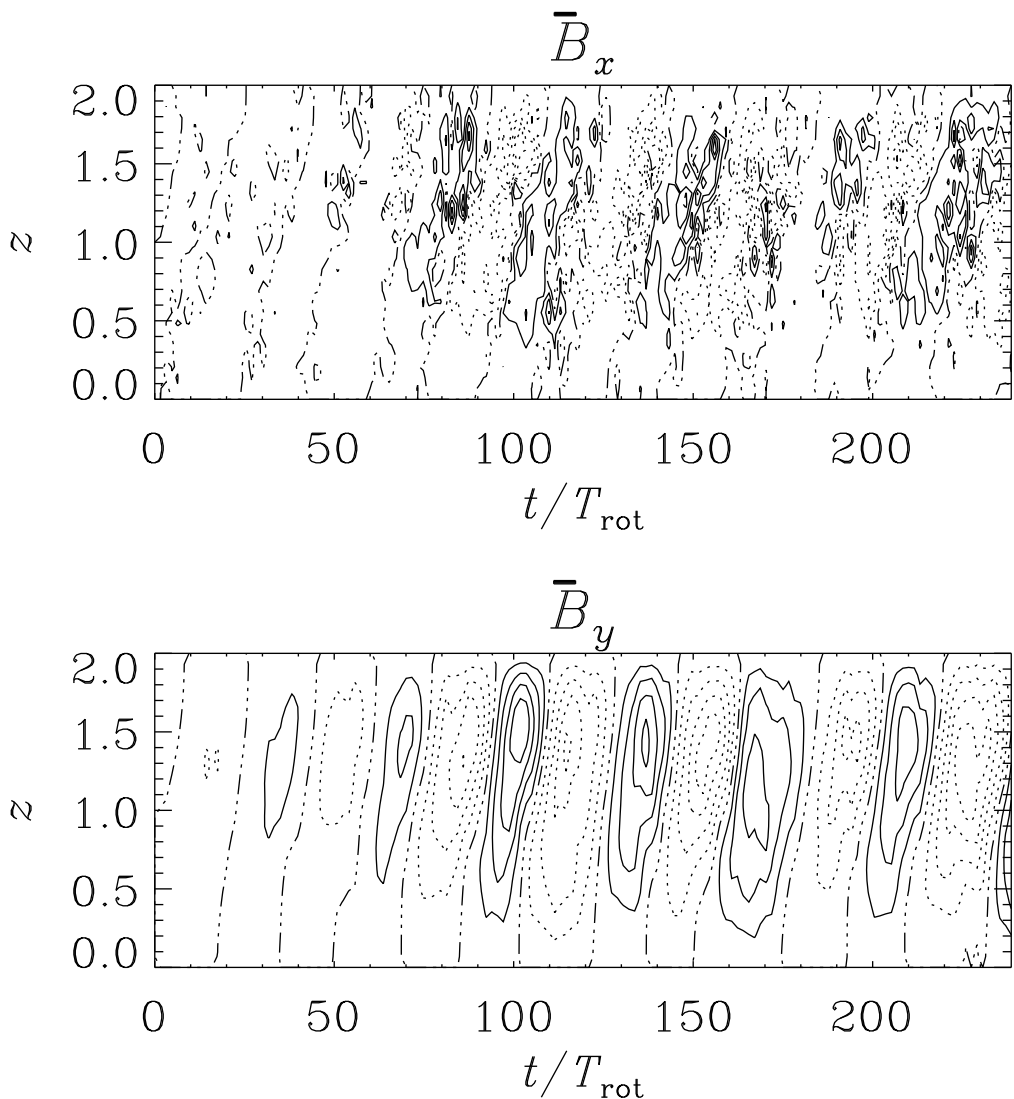}
\caption[]{
Mean-field calculation with $\alpha_N=0.01$, $\alpha_0=-0.001$,
and $\eta_0=0.005$.
}\label{Fpp_alpn=0.01}\end{figure}

Given the high level of noise found in the transport coefficients it is
useful to elaborate further on this point. In the following we present
a model calculation where the effect of strong noise is included. We
solve \Eqss{dyneq1}{dyneq2} and replace \Eq{chosen} by
\EQ
\alpha_{yy}=-0.001\Omega z+\alpha_{\rm N}N,\quad\eta_{\rm t}=0.0055\Omega H^2,
\label{chosen2}
\EN
where $\alpha_{\rm N}$ is a coefficient and $N(z,t)$ is gaussian
noise in $z$ and $t$, normalized to a root-mean-square value of unity.
The other components of $\alpha_{ij}$ and $\eta_{ij}^*$ are ignored
because we want to examine the effect of noise and take therefore
the simplest possible model that was presented already in \Fig{Fmodel}.
In \Fig{Fpp_alpn=0.01} we present the result for $\alpha_{\rm N}=0.01$,
corresponding to a noise level that exceeds the mean alpha-effect by a
factor ten or more, especially close to the midplane, where the coherent
alpha-effect vanishes.

\epsfxsize=16cm

The resulting field structure looks rather similar to the mean field
obtained by averaging the data from the original three-dimensional
simulations. The noise level exceeds the mean (coherent) alpha-effect by
more than a factor of ten, which is comparable to or even larger than
what has been suggested in connection with the solar dynamo (Choudhuri
1992, Moss et al.\ 1992, Hoyng et al.\ 1994, Otmianowska-Mazur et al.\
1997). The main point we want to make here is that a high noise level in
the mean-field transport coefficients (\Sec{fitting}) is quite natural
and not simply an indication of inaccurate measurements.

\section{Comments on nonlinear quenching}

The issue of magnetic quenching of turbulent transport coefficients has
not been addressed in the present paper. One must expect that all
components of the $\alpha$ and $\eta$ tensors
depend on magnetic field strength and field structure. In
principle our approach could be modified to allow for such nonlinearities. The
usual assumption is that, if the turbulent transport coefficients are
caused by turbulent convection, for example, $\alpha$ and $\eta_{\rm t}$ would
decrease with increasing field strength. On the other hand, if the
turbulent transport coefficients are caused by flow fields which themselves
are magnetically driven, for example by magnetic (e.g., Balbus-Hawley
and Parker) instabilities, then $\alpha$ and $\eta_{\rm t}$ may actually increase
with increasing field strength (see Brandenburg 1997 for a recent
discussion on that). There is even some observational evidence for this
somewhat unusual proposal (Brandenburg, Saar \& Turpin 1998,
Saar \& Brandenburg 1999).
Thus, in addition to allowing $\alpha$ and $\eta_{\rm t}$ to be scale
dependent, one should also allow for the possibility that $\alpha$ and
$\eta_{\rm t}$ may actually increase with magnetic field strength.
However, these type of behaviors must still be compatible with helicity
conservation.

In the case of a fully periodic
domain the situation is now fairly well understood. For example,
we know that the correct helicity limited growth of the large scale field
can be reproduced if both $\alpha$ and $\eta_{\rm t}$ are equally strongly
quenched by the mean magnetic field, $\meanBB$, and the quenching is of the form
$1/(1+\alpha_B\meanBB^2/B_{\rm eq}^2)$, where $B_{\rm eq}$
is the equipartition field strength, and the coefficient $\alpha_B$ is
proportional to the magnetic Reynolds number. This type of `catastrophic'
quenching goes back to the early paper by Vainshtein \& Cattaneo (1992),
and was confirmed using simulations (Cattaneo \& Hughes 1996),
but it is now clear that this behavior is primarily a consequence of
magnetic helicity conservation (Blackman \& Field 2000, B01), and
independent of the type of nonlinearity (Brandenburg \& Subramanian 2000).
Evidence for `catastrophic' $\eta_{\rm t}$-quenching was first
found by Cattaneo \& Vainshtein (1991) using two-dimensional simulations.
Three-dimensional simulations of isotropic helical turbulence in a
box yielded a resistively slow saturation behavior of $\alpha$ which is
well described by a fit of the form $\meanBB^2\sim1-\exp(-2\eta k^2\Delta t)$, where $\eta$
is the microscopic magnetic diffusivity, $k$ is the wavenumber of the
large scale field, and $\Delta t$ is the time after which the small
scale field has saturated. The necessity to obey this fit is
sufficiently stringent to exclude for example cubic quenching behavior
(e.g., Moffatt 1972, R\"udiger 1974, R\"udiger \& Kitchatinov 1993),
or quenching of the form $1-\alpha_B\meanBB^2\!/B_{\rm eq}^2$.
As was recently pointed out by Field \& Blackman (2002),
a type of quenching that satisfies the magnetic helicity equation
exactly and that is therefore also compatible with resistively limited
growth is dynamical quenching (Kleeorin, Rogachevskii \& Ruzmaikin 1995).
According to this description, $\alpha$ and $\eta_{\rm t}$ are no longer
catastrophically quenched. In other words, the earlier conclusion of B01
is not compulsory because of a degeneracy of the dynamo equations in
the fully helical case.

In a nonperiodic domain the situation is quite different,
because there is then the possibility of a magnetic helicity flux out
of the domain. This magnetic helicity loss is associated with a loss
of magnetic energy. In the model of BD much of the field that was lost through
the boundaries was of large scale, which made the large scale dynamo
process less efficient. This corresponds to a less drastic quenching,
because $\alpha_B$ is now only proportional to the square root of the
magnetic Reynolds number (BD).

The presence of shear affects the saturation results in a different
way. Field amplification by shear is quite independent of the
magnetic helicity effect, but this concerns only the toroidal field
amplification. For the dynamo to work one still needs to generate poloidal
field in order to make use of the shear. Thus, saturation of the large
scale field still occurs on a resistive time scale, but larger field
amplitudes are now possible (Brandenburg, Bigazzi \& Subramanian 2001)
Because of shear, the field evolves in an oscillatory fashion
(as predicted by mean-field $\alpha\Omega$-dynamo theory; e.g. Moffatt
1978). It is at present not entirely clear, however, whether the time scale for
this oscillation is resistive or dynamical.

\section{Conclusions}

In the present paper we have demonstrated that the results of the
three-dimensional numerical nonlinear dynamo simulations of BNST95 can
be fitted by a mean-field dynamo model with properly chosen
parameterizations of transport coefficients. The parameterization found
from such a fitting procedure is
far from that of naive kinematic mean-field dynamo theory. In fact, our
investigations point toward the possibility that turbulent transport
coefficients might be wavenumber dependent. This possibility was first
explored by Moffatt (1983), who used a renormalization group approach
to obtain differential equations for $\alpha$ and $\eta_{\rm t}$ as functions
of wavenumber. We find that the strongest contribution to $\alpha$ and
$\eta_{\rm t}$ tends to come from the largest scales in the system.
Unfortunately the scatter in the data is considerable and only
data for the first few wavenumbers seem to be significant. However,
simple representations of the form \eq{approx_alp} and \eq{approx_eta}
seem to reproduce the original data reasonably well (compare
Fig.~\ref{Fnonloc} with Fig.~\ref{Fbutter}). These representations are
consistent with fits to the electromotive force and, more importantly,
the corresponding energy dissipation is now positive.

Note that we have restricted the diffusion tensor that multiplies the
current to be diagonal. This restriction means that the turbulent
diffusion does not mix $x$ and $y$ components of the mean magnetic
field. Such kind of `cross-talk' is absent in dynamo models with
short-correlated random velocity fields (Molchanov et al.\ 1985),
although it may exist in other models (R\"adler's $\Omega\times
J$-effect is an example; see Krause \& R\"adler 1980). We note that our
estimates for the $\alpha$-tensor are not significantly modified when
the diffusion tensor is restricted to be diagonal.

Comparing the results of \Secs{Scomparison}{Slocal}--\ref{Snonlocalstar},
several similar aspects can be recovered in each case. Firstly,
$\alpha_{yy}$ is negative in the upper hemisphere (or disc plane) and
its magnitude, in natural units, is around $10^{-3}$ at $z\approx H$
(see \Figs{Fpalp_loc}{Fpalp_locJ}) or at small wave numbers ($k=k_1$),
i.e.\ at large scales (see \Figs{Fpalp}{FpalpJ}). Secondly, the turbulent
magnetic diffusivity in the streamwise direction ($y$) is such that
the corresponding mean-field dynamo is marginally excited. The
turbulent magnetic diffusivity in the cross-stream direction ($x$) is
however much smaller than in the streamwise direction (cf.\ the
two panels in \Fig{Fpeta_locJ}). This is again reflected in the
nonlocal approach (cf.\ the two panels in \Fig{FpetaJ}), but
here the diffusion coefficient of $B_x$ is more noisy and, at the
minimum wavenumber $k_0$, the diffusion coefficient of $B_x$
becomes comparable to that of $B_y$, but it stays at least positive.
We regard similarities in the results of the various approaches as
an indication of their robustness (e.g., the values of $\alpha_{yy}$
and $\eta_{xx}^*$). On the other hand, dissimilarities and noisy results,
as well as large error bars, indicate that results are not reliable
(e.g.\ $\eta_{yy}^*$ and $\hat\eta_{yy}^*$). Positivity of the
diffusion coefficients is an additional criterion for reliability.

We should stress that we expect realistic transport coefficients to
show both height dependence as well as scale dependence. In addition,
the $\alpha$-effect is a dynamical one that satisfies an explicitly
time-dependent equation (Rogachevskii \& Kleeorin 2001). It is however
rather difficult to determine all these aspects simultaneously, which
is why we have looked at each of them separately. Indeed, our goal is
therefore not to distinguish between the various approaches, but rather
to present a first assessment of the presence of each of these effects
(height dependence, scale dependence), all of which should be present
simultaneously.

We stress that our final model is based on very noisy fits for
$\alpha_{yy}$. We feel that it is important to acknowledge the reality
and physical significance of unsteady and random transport coefficients
and to investigate models with stochastic alpha. Note also that there
are some models showing the possibility of the generation of large
scale magnetic patterns in a flow with a random (incoherent) $\alpha$-effect
(Vishniac \& Brandenburg 1997). However, in order to reproduce the butterfly
(space-time) diagram obtained from the three-dimensional simulations a
coherent (non-noisy) $\alpha$-effect must still be present. The present
results suggest that the simulations are reproduced with a coherent
$\alpha$-effect that is only about 10\% of the incoherent $\alpha$-effect
in parts of the domain.

The idea that only the smallest wavenumbers contribute to the transport
coefficients may well prove to be a reasonable representation for
astrophysical dynamos. Already now there is some evidence for this
proposal: so far only models that include the lowest wavenumber are
able to reproduce stellar cycle frequencies that decrease sufficiently
rapidly with increasing rotational frequency (Brandenburg, Saar \&
Turpin 1998). By contrast, models with transport coefficients that are
independent of wavenumber (Tobias 1998) show that the cycle frequency
decreases too slowly with increasing rotational frequency.

Although the present work was motivated by astrophysical and geophysical
applications, the analysis presented here may just as well apply to
homogeneous dynamos in the laboratory such as the Karlsruhe experiment
(Stieglitz \& M\"uller 2001). The results from these experiments can
now accurately be tested against simulations (Tilgner 2000) and mean-field
theory (R\"adler \ea 1998). Due to the presence of boundaries, however,
the $\alpha$ and $\eta$ tensors have to be replaced by integral kernels
(Stefani, Gerbeth \& R\"adler 2000). One may therefore expect that the
approach explored in the present paper to determine integral kernels for
the $\alpha$ and $\eta$ tensors will soon gain in importance.

\section*{Acknowledgements}
We thank David Moss and G\"unther R\"udiger for helpful suggestions that
have helped to improve the presentation of the paper.
One of us (DS) is grateful for financial support by the Russian Foundation for
Basic Research under grant 99-01-00362, the NATO Collaborative Research Grant
PST.CLG 974737, as well as the Royal Society for an ex-quota grant,
which enabled him to spend half a year in Newcastle, where most of this
work was carried out.

\section*{References}

\begin{list}{}{\leftmargin 3em \itemindent -3em\listparindent \itemindent
\itemsep 0pt \parsep 1pt}\item[]

Arlt, R., \& Brandenburg, A.\yana{2001}{380}{359}
{372}{Search for non-helical disc dynamos in simulations}

Balbus, S. A. \& Hawley, J. F.\yapj{1991}{376}{214}
{222}{A powerful local shear instability in weakly magnetized disks.
I. Linear analysis}

Blackman, E. G. \& Field, G. F.\yapj{2000}{534}{984}
{988}{Constraints on the magnitude of $\alpha$ in dynamo theory}

Brandenburg, A.\yjour{1997}{Acta Astron.\ Geophys.\ Univ.\ Comenianae}{XIX}{235}
{261}{Large scale turbulent dynamos}

Brandenburg, A.\yproc{1998}{61}
{86}{Disc Turbulence and Viscosity}
{Theory of Black Hole Accretion Discs}
{M. A. Abramowicz, G. Bj\"ornsson \& J. E. Pringle}
{Cambridge University Press}

Brandenburg, A.\yapj{2001}{550}{824}
{840}{The inverse cascade and nonlinear alpha-effect in simulations
of isotropic helical hydromagnetic turbulence} (B01).

Brandenburg, A., Bigazzi, A., \& Subramanian, K.\ymn{2001}{325}{685}
{692}{The helicity constraint in turbulent dynamos with shear}

Brandenburg, A., \& Dobler, W.\yana{2001}{369}{329}
{338}{Large scale dynamos with helicity loss through boundaries}
(BD).

Brandenburg, A. \& Donner, K. J.\ymn{1997}{288}{L29}
{L33}{The dependence of the dynamo alpha on vorticity}

Brandenburg, A., \& Sarson, G. R.\yprl{2002}{88}{055003-1}
{4}{The effect of hyperdiffusivity on turbulent dynamos with helicity}

Brandenburg, A., Nordlund, \AA., Pulkkinen, P.,
Stein, R.F., \& Tuominen, I.\yana{1990}{232}{277}
{291}{3-D Simulation of turbulent cyclonic magneto-convection}

Brandenburg, A., Nordlund, \AA., Stein, R. F.
\& Torkelsson, U.\yapj{1995}{446}{741}
{754}{Dynamo generated turbulence and large scale magnetic fields
in a Keplerian shear flow} (BNST95).

Brandenburg, A., Nordlund, \AA., Stein, R. F.
\& Torkelsson, U.\yapjl{1996}{458}{L45}
{L48}{The disk accretion rate for dynamo generated turbulence} (BNST96).

Brandenburg, A., Saar, S. H. \& Turpin, C. R.\yapjl{1998}{498}{L51}
{L54}{Time evolution of the magnetic activity cycle period}

Brandenburg, A. \& Subramanian, K.\yana{2000}{361}{L33}
{L36}{Large scale dynamos with ambipolar diffusion nonlinearity}

Cattaneo, F., \& Vainshtein, S. I.\yapjl{1991}{376}{L21}
{L24}{Suppression of turbulent transport by a weak magnetic field}

Cattaneo, F., \& Hughes, D. W.\ypr{1996}{E 54}{R4532}
{R4535}{Nonlinear saturation of the turbulent alpha effect}

Choudhuri, A. R.\yana{1992}{253}{277}
{285}{Stochastic fluctuations of the solar dynamo}

Dittrich, P., Molchanov, S. A., Sokoloff, D. D. \& Ruzmaikin, A. A.\yan{1984}
{305}{119}{125}{Mean magnetic field in renovating random flow}

Elperin, T., Kleeorin, N., Rogachevskii, I. \&
Sokoloff, D.\ypr{2000}{E 61}{2617}
{2625}{Passive scalar transport in a random flow with a finite renewal
time: Mean-field equation}

Field, G. B., \& Blackman, E. G.\yapj{2002}{572}{685}
{692}{Dynamical quenching of the $\alpha^2$ dynamo}

Glatzmaier, G. A. \& Roberts, P. H.\ynat{1995}{377}{203}
{209}{A three-dimensional self-consistent computer simulation
of a geomagnetic field reversal}

Gruzinov, A. V. \& Diamond, P. H.\yprl{1994}{72}{1651}
{1653}{Self-consistent theory of mean-field electrodynamics}

Hasler, K.-H., Kaisig, M. \& R\"udiger, G.\yana{1995}{295}{245}
{248}{Diffusion approximation probed with Parker instability simulations}

Hawley, J. F., Gammie, C. F. \& Balbus, S. A.\yapj{1995}{440}{742}
{763}{Local three-dimensional magnetohydrodynamic simulations
of accretion discs}

Hawley, J. F., Gammie, C. F. \& Balbus, S. A.\yapj{1996}{464}{690}
{703}{Local three dimensional simulations of an accretion disk
hydromagnetic dynamo}

Hollerbach, R., Barenghi, C. F. \& Jones, C. A.\ygafd{1992}{67}{3}
{25}{Taylor constraint in a spherical alpha-omega-dynamo}

Hoyng, P., Schmitt, D. \& Teuben, L. J. W.\yana{1994}{289}{265}
{278}{The effect of random alpha-fluctuations and the global properties
of the solar magnetic field}

Jones, C. A. \& Wallace, S. G.\ygafd{1992}{67}{37}
{64}{Periodic, chaotic and steady solutions in alpha-omega-dynamos}

Keinigs, R. K.\ypf{1983}{26}{2558}
{2560}{A new interpretation of the alpha effect}

Kleeorin, N. I, Rogachevskii, I., \& Ruzmaikin, A.\yana{1995}{297}{159}
{167}{Magnitude of the dynamo-generated magnetic field in solar-type
convective zones}

Krause, F. \& R\"adler, K.-H.\ybook{1980}
{Mean-Field Magnetohydrodynamics and Dynamo Theory}
{Akademie-Verlag, Berlin; also Pergamon Press, Oxford}

Miller, K.\ A. \& Stone, J.\ M.\yapj{2000}{534}{398}
{419}{The formation and structure of a strongly magnetized corona above 
a weakly magnetized accretion disk}

Moffatt, H. K.\yjfm{1972}{53}{385}
{399}{An approach to a dynamic theory of dynamo action
in a rotating conducting fluid}

Moffatt, H. K.\ybook{1978}
{Magnetic Field Generation in Electrically Conducting Fluids}
{Cambridge University Press, Cambridge}

Moffatt, H. K.\yjour{1983}{Rep.\ Prog.\ Phys.}{46}{621}
{664}{Transport effects associated with turbulence with particular
attention to the influence of helicity}

Molchanov, S. A., Ruzmaikin, A. A. \& Sokoloff, D. D., ``Kinematic 
dynamo in random flow," {\it Sov. Phys. Usp.} {\bf 28}, 307 -- 327 
(1985).

Moss, D., Brandenburg, A., Tavakol, R. K. \& Tuominen, I.\yana{1992}{265}{843}
{849}{Stochastic effects in mean field dynamos}

Nicklaus, B. \& Stix, M.\ygafd{1988}{43}{149}
{166}{Corrections to first order smoothing in mean-field electrodynamics}

Ossendrijver, M., Stix, M., \& Brandenburg, A.\yana{2001}{376}{713}
{726}{Magnetoconvection and dynamo coefficients: dependence of the
$\alpha$-effect on rotation and magnetic field}

Otmianowska-Mazur, K.\ygafd{1997}{86}{229}
{247}{The turbulent EMF as a time series and the 'quality' of dynamo cycles}

Parker, E. N.\yapj{1955}{122}{293}
{314}{Hydromagnetic dynamo models}

Parker, E. N.\ybook{1979}{Cosmical Magnetic Fields}{Clarendon Press, Oxford}

Piddington, J. H.\ysph{1972}{22}{3}
{19}{Solar dynamo theory and the models of Babcock and Leighton}

R\"adler, K.-H., Apstein, E., Rheinhardt, M., Sch\"uler, M.\yjour{1998}
{Studia Geophys.\ et Geod.\ }{42}{224}
{231}{The Karlsruhe dynamo experiment. A mean field approach}

Reg\"os, E.\ymn{1997}{286}{104}
{114}{Parker's instability and viscosity in quiescent cataclysmic variable discs}

Roberts, P. H. \& Soward, A. M.\yan{1975}{296}{49}
{64}{A unified approach to mean field electrodynamics}

Rogachevskii, I. \& Kleeorin, N.\ypr{2001}{E 64}{056307}
{1-14}{Nonlinear turbulent magnetic diffusion and mean-field dynamo}

R\"udiger, G.\yan{1974}{295}{275}
{284}{The influence of a uniform magnetic field of arbitrary strength
on turbulence}

R\"udiger, G. \& Kitchatinov, L. L.\yana{1993}{269}{581}
{588}{Alpha-effect and alpha-quenching}

R\"udiger, G. \& Pipin, V. V.\yana{2000}{362}{756}
{761}{Viscosity-alpha and dynamo-alpha for magnetically driven
compressible turbulence in Kepler disks}

Saar, S. H. \& Brandenburg, A.\yapj{1999}{524}{295}
{310}{Time evolution of the magnetic activity cycle period. II.
Results for an expanded stellar sample}

Steenbeck, M., Krause, F. \& R\"adler, K.-H.\yjour{1966}
{Z. Naturforsch.}{21a}{369}
{376}{Berechnung der mittleren Lorentz-Feldst\"arke $\overline{\vv\times\BB}$
f\"ur ein elektrisch leitendendes Medium in turbulenter, durch
Coriolis-Kr\"afte beeinflu{\ss}ter Bewegung}
See also the translation in Roberts \& Stix, The turbulent dynamo...,
Tech. Note 60, NCAR, Boulder, Colorado (1971).

Stefani, F., Gerbeth, G., \& R\"adler, K.-H.\yan{2000}{321}{65}
{73}{Steady dynamos in finite domains: an integral equation approach}

Stieglitz, R., \& M\"uller, U.\ypf{2001}{13}{561}
{564}{Experimental demonstration of a homogeneous two-scale dynamo}

Tilgner, A.\yjour{2000}{Phys. Earth Planet. Int.}{117}{171}
{177}{Towards experimental fluid dynamos}

Tobias, S.\ymn{1998}{296}{653}
{661}{Relating stellar cycle periods to dynamo calculations}

Tout, C. A. \& Pringle, J. E.\ymn{1992}{259}{604}
{612}{Accretion disc viscosity: A simple model for a magnetic dynamo}

Vainshtein, S. I. \& Cattaneo, F.\yapj{1992}{393}{165}
{171}{Nonlinear restrictions on dynamo action}

Vishniac, E. T. \& Brandenburg, A.\yapj{1997}{475}{263}
{274}{An incoherent $\alpha-\Omega$ dynamo in accretion disks}

Vishniac, E. T. \& Cho, J.\yapj{2001}{550}{752}
{760}{Magnetic helicity conservation and astrophysical dynamos}

Zheligovsky, V. A., Podvigina, O. M., \& Frisch, U.\pgafd{2001}
{Dynamo effect in parity-invariant flow with large and moderate
separation of scales}
{\sf nlin.CD/0012005}.

Ziegler, U., \& R\"udiger, G.\yana{2000}{356}{1141}
{1148}{Angular momentum transport and dynamo-effect in stratified,
weakly magnetic disks}

\end{list}
\end{document}